\documentclass[aps,prd,superscriptaddress]{revtex4-2}

\usepackage{amsmath}
\usepackage{anysize}
\usepackage{comment}
\PassOptionsToPackage{hyphens}{url}\usepackage[pdftex]{hyperref}
\usepackage{float}
\usepackage{latexsym}
\usepackage{lineno}
\usepackage{graphicx}
\usepackage{comment} 
\usepackage{xcolor}
\usepackage{todonotes}

\advance\textheight1cm
\advance\voffset-1cm
\usepackage[]{hyperref}
\hypersetup{
colorlinks,
linkcolor={blue},
linktocpage=true,
citecolor={red},
urlcolor={olive}
}

\newcommand{\mue}{$\mu-e$}

\newcommand{\MuToEmConv}{$\mu^{-}+N\rightarrow e^{-} +N$}
\newcommand{\MuToEpConv}{$\mu^{-}+N\rightarrow e^{+} +N'$}
\newcommand{\MuToEX}{$\mu^{-}\rightarrow e^{-} + X$}
\begin{document}

\title{ Experimental Searches for Muon to Electron Conversion in a Nucleus: COMET, DeeMe and Mu2e \\ Contributed paper for Snowmass 21 \\}

\author{M.J.~Lee }
\affiliation{for the COMET Collaboration, Sungkyunkwan University, Suwon, GyeongGi-do, 16419, Republic of Korea}

\author{S.~Middleton}
\affiliation{for the Mu2e Collaboration, California Institute of Technology, 1200 E California Blvd, Pasadena, CA 91125, USA}

\author{Y.~Seiya}
\affiliation{for the DeeMe Collaboration, Osaka City University, 3-3-138 Sugimoto, Sumiyoshi, Osaka, 558-8585, Japan}


\today

FERMILAB-FN-1150

\begin{abstract}

   Searches for charged lepton flavor violation (CLFV) offer deep probes for a range of new physics scenarios, such as super-symmetric models, theories involving scalar leptoquarks or additional Higgs doublets, and models explaining the neutrino mass hierarchy and the matter-antimatter asymmetry of the universe via leptogenesis. The coherent, neutrinoless conversion of a muon to an electron in the field of a nucleus, $\mu^{-}+N\rightarrow e^{-} +N$, is one example of a muonic CLFV process which has sensitivity to this new physics. This paper details three experiments: COMET, DeeMe and Mu2e which will search for $\mu^{-}+N\rightarrow e^{-} +N$ in the coming decade.  These experiments offer sensitivity up to an effective new physics mass scale of $\mathcal{O}(10^{4}$ TeV/c$^{2}$), going far beyond what can be achieved in direct, collider-based, searches. The theoretical motivation, designs and anticipated timelines for these three experiments are presented. These experiments are a crucial part of a global search for CLFV. Continued support for all planned experimental searches for muonic CLFV is strongly encouraged.
    
\end{abstract}

\maketitle

\tableofcontents

\newpage

\section{Executive Summary}
\begin{itemize}

\item Searches for charged lepton flavor violation (CLFV) at the Intensity Frontier provide deep probes for new physics models and are complementary to Energy Frontier searches for new physics.

\item CLFV processes offer unique discovery potential for many beyond the Standard Model (BSM) scenarios. Observation of a CLFV transition would be an undeniable sign of new physics, which goes beyond that induced by non-zero neutrino masses. 

\item Historically, muons have provided the best experimental sensitivity to CLFV. This is a result of the muon's properties, including its lifetime. A huge number of muons can be produced from tertiary beams at several facilities around the globe. In the near future, new muon beamlines will be constructed, leading to increases in beam intensity by several orders of magnitude.

\item Several experiments will come online in the next few years including: DeeMe, COMET and Mu2e, which search for \MuToEmConv~, MEG-II, searching for  $\mu^{+} \rightarrow e^{+} + \gamma$, and Mu3e, searching for $\mu^{+} \rightarrow e^{+} + e^{-} + e^{+}$. These muon sector searches are an integral part of a global search for CLFV. Searches for CLFV in the tau sector and lepton flavor violating Higgs decays will also take place at collider based experiments.

\item New physics contributions to CLFV in dimension-6 operators can be divided into dipole, or photonic contributions, where the CLFV is mediated by a photon that is absorbed by the nucleus, and contact, or four-fermion, interactions. Searches for $\mu^{+} \rightarrow e^{+} + \gamma$ are sensitive to only photonic processes and searches for \MuToEmConv~ or $\mu^{+} \rightarrow e^{+} + e^{-} + e^{+}$  are sensitive to photonic and four fermion (contact) processes. 
Mu2e and COMET Phase-II can reach to effective new physics mass scales of up to $\mathcal{O}(10^{4}$ TeV/c$^{2}$) \cite{deGouvea2013}.

\item This paper details the three upcoming experimental searches for the coherent neutrinoless conversion of a muon to an electron in the field of a nucleus, \MuToEmConv.

\item The current upper limit of the rate of  \MuToEmConv~ relative to ordinary muon capture, $R_{\mu e}$, was set by the SINDRUM-II experiment at $ 7 \times 10^{-13} $ ($ 90 \% CL$) using a gold target. The Mu2e experiment, under construction at Fermilab, aims to improve on this limit down to $ \mathcal{O}(10^{-17})$.  The  COMET  collaboration  is  preparing  their  apparatus  at  J-PARC,  with  planned  sensitivity  of $\mathcal{O}(10^{-15})$ for Phase-I and $\mathcal{O}(10^{-17})$ for Phase-II.  Both Mu2e and COMET will use an aluminum target. The DeeMe experiment, also based at J-PARC, is preparing for a sensitivity of $\mathcal{O}(10^{-13})$ using a carbon target.

\item In addition, to the primary \MuToEmConv~ search, COMET and Mu2e have improved sensitivity to \MuToEpConv~, which also violates lepton number ($\Delta L = 2$), and the exotic \MuToEX~ process where
$X$ can be an axion-like particle (ALP), familon, or majoron with a lepton
flavor violating coupling to leptons.


\item We urge for the strong, continued support for all experimental searches for CLFV in the muon sector. US participation in searching for CLFV
in the muon sector is critical to the realization of the program's physics
goals.

\end{itemize}

\clearpage

\section{Introduction}

\label{introduction}

\subsection{Motivations}
Unlike for the quarks and neutral leptons (neutrinos), flavor violation has never been observed for the charged leptons. Physicists began searching for charged lepton flavor violation (CLFV) in  $\mu \rightarrow e \gamma$ as soon as the muon was discovered; the search continues to this day, along with searches for other charged lepton flavor violating processes. We know CLFV must occur because neutrinos oscillate, but the calculated rates are $\mathcal{O}(10^{-54})$, far beyond the reach of any conceivable experiment \cite{Marciano2008}. Therefore, any experimental observation of a CLFV process would be a true departure from the Standard Model (SM), even when minimally extended to include massive neutrinos,
and would conclusively demonstrate the presence of beyond Standard Model (BSM) physics. Many well-motivated extensions of the SM predict much higher rates of the CLFV processes which fall within the reach of the next generation of CLFV experiments: COMET \cite{ 10.1093/ptep/ptz125}, DeeMee \cite{DeeMe2010}, Mu2e \cite{mu2eTDR} (all searching for \MuToEmConv~); MEG-II \cite{MEGII2018} (searching for  $\mu^{+} \rightarrow e^{+} + \gamma$); and  Mu3e \cite{Mu3e2013} (searching for $\mu^{+} \rightarrow e^{+} e^{-} e^{+}$ ). Examples of such new physics include SO(10) super-symmetry \cite{Calibbi2014,Calibbi2006}, models of scalar lepto-quarks \cite{Arnold, heeck} and models with additional Higgs doublets \cite{Abe2017}. In addition, searches for \MuToEmConv~ can also help place limits on lepton flavor violating Higgs decays, Ref.~\cite{harnik} suggests that Mu2e and COMET have sensitivity up to $BR (h \rightarrow \mu e)$ of $10^{-10}$. CLFV measurements can also help elucidate the mechanism behind neutrino masses.  Since different mass generating Lagrangians predict different rates of CLFV, determining the rates of multiple CLFV processes allows us to discriminate among models, Ref.~\cite{HAMBYE201413}

This paper describes the design and physics potential of three upcoming searches for \MuToEmConv~ in Sections~\ref{DeeMe}~(DeeMe), \ref{COMET}~(COMET), and \ref{mu2e}~(Mu2e). Direct searches, such as those carried out at the LHC, have
not produced conclusive evidence for BSM physics. Searches for rare CLFV processes at the Intensity Frontier, such as those presented in this paper, can indirectly probe BSM regimes far beyond the reach of direct searches at the Energy Frontier. For example, Mu2e and COMET Phase-II can reach up to an effective new physics mass scale of $\mathcal{O}(10^{4}$ TeV/c$^{2}$) \cite{deGouvea2013}. 

Muons have consistently provided powerful constraints on CLFV since their discovery. Ref. \cite{bernstein_history} details historical searches for CLFV. This is a consequence of the properties of the muon itself. Muons have a relatively long lifetime of 2.2 $\mu$s and a mass which provides only a few decay channels, resulting in much simpler final states than those of the heavier tau lepton. In addition, intense muon beams can be produced as by-products from high energy protons on target at facilities such as J-PARC and Fermilab, providing the high statistics required to investigate processes with very small rates. 
 New physics contributions to CLFV in dimension-6 operators can be divided into dipole, or photonic contributions, where the CLFV is mediated by a photon that is absorbed by the nucleus, and contact interactions, where a new virtual particle is exchanged which couples to the fermions. Searches for radiative $\mu^{+} \rightarrow e^{+} + \gamma$ decays are only sensitive to electromagnetic dipole interactions. On the other hand, experiments searching for \MuToEmConv~ and $\mu^{+} \rightarrow e^{+} e^{-} e^{+}$ are sensitive to conversion process that result from either photonic or four fermion (contact) type interactions. 
 While searches for $\mu^{+} \rightarrow e^{+} + \gamma$ provide powerful constraints on the dipole operators, \MuToEmConv~ conversion is the most sensitive observable to explore operators involving quarks \cite{deGouvea2013}. 
 
The relative contributions to the different CLFV channels reflects the nature of the underlying physics model responsible \cite{calibbi2017charged}. It is, therefore, crucial to explore all three muon CLFV channels to understand the potential signature of new physics. In the event that CLFV signals are observed in the next generation of experiments, comparisons among the results from experiments dedicated to different CLFV channels can help elucidate the nature of new physics responsible. Thus, searches for \MuToEmConv~ are a crucial part of an active global CLFV program and are complementary to the searches at MEG-II and Mu3e.  In addition, the rate of conversion is dependent on the atomic nucleus \cite{kitano2002}. Therefore, another way to elucidate a conversion signal would be to measure the conversion rate in a material complementary to aluminum (the target material proposed in Mu2e and COMET) or carbon (the target material proposed for DeeMe). Titanium is one proposed alternative, but as seen in Ref.~\cite{kitano2002}, splitting among the rates for different operators becomes larger as $Z$ increases. Hence taking the relative rate in two complementary materials may suggest what type of physics (vector, scalar or dipole) is responsible.

\subsection{Neutrinoless Muon to Electron Conversion}
 
 As described in the preceding section, the process of coherent, neutrinoless muon to electron conversion  in the nuclear field, \MuToEmConv,
 probes a wide spectrum of new physics models. The rate of this conversion relative to ordinary muon capture is conventionally defined as:

\begin{equation}
R_{\mu e} = \frac{\Gamma(\mu^{-} + N(A,Z) \rightarrow e^{-} + N(A,Z))}{\Gamma(\mu^{-} + N(A,Z) \rightarrow \text{all captures})},
\end{equation}
where $N(A,Z)$ denotes the mass and atomic numbers of the target nuclei. The present experimental limit on $R_{\mu e}$ is $7 \times 10^{-13}$ $(90\% CL)$ set by the SINDRUM-II experiment using a gold target \cite{Bertl2006}. The Mu2e experiment \cite{mu2eTDR}, under construction at Fermilab, aims reach a single event sensitivity (SES) of $\sim 3 \times 10^{-17}$ on the conversion rate, a 90$\%$ CL on $R_{\mu e}$ of $6 \times 10^{-17}$, and a $5\sigma$ discovery potential at $\sim 2 \times (10^{-16})$. The COMET collaboration is preparing their apparatus at J-PARC, with planned sensitivity of $\mathcal{O}(10^{-15})$ for Phase-I \cite{10.1093/ptep/ptz125} and a similar sensitivity to Mu2e for Phase-II~\cite{Lee2018}. Both Mu2e and COMET will use an aluminum target. The DeeMe experiment, also based at J-PARC, is preparing for a sensitivity of $\sim \mathcal{O}
(10^{-13})$ on carbon ~\cite{Teshima2019}. An extension to the Mu2e physics program, Mu2e-II \cite{Mu2eII} has also been proposed and would reach a sensitivity of $\mathcal{O}(10^{-18})$, but Mu2e-II will not be discussed in this paper. In addition, MEG-II has a projected sensitivity  on $\mu^{+} \rightarrow e^{+} \gamma$ down to $6 \times 10^{-14}$, an order of magnitude better than the current upper limit  \cite{MeGLimit}. Mu3e has a projected sensitivity  on $\mu^{+} \rightarrow e^{-} e^{+} e^{+} $ of $10^{-15}-10^{-16}$ several orders of magnitude better than the current upper limit, $BR(\mu^{+} \rightarrow e^{-} e^{+} e^{+})< 1 \times 10^{-12}$ \cite{Mu3eLimit}.  
 
 When  negatively charged muons are trapped in the field of the target nucleus, forming a muonic atom, and the muon cascades down in energy to the muonic $1s$ bound state there are three possibilities considered:
 
 \begin{enumerate}
     \item \textbf{decay in orbit (DIO)} $\mu^{-} \rightarrow \nu_{\mu} + \bar{\nu}_{e} + e^{-}$, a background which must be eliminated;
     \item \textbf{muon capture} $\mu^{-} + N(A,Z)\rightarrow \text{all captures}$, which can be used to normalize the signal rate ($R_{\mu e}$);
     \item \textbf{neutrinoless conversion} $\mu^{-} + N(A,Z)\rightarrow e^{-} + N(A,Z)$, the conversion signal.
 \end{enumerate}
 
 The relative rates of capture and decay are nucleus dependent: in aluminum the capture rate is 61$\%$ and the decay rate is 39$\%$, whereas in carbon the capture rate is just 8 $\%$ and the decay rate is 92$\%$.
 
 \MuToEmConv~ in a muonic atom results in the emission of a mono-energetic electron with an energy:
 
 \begin{equation}
     E_{\mu e} = m_{\mu} - E_{BE,1s} - E_{recoil},
 \end{equation}
 where $m_{\mu}$ (105.66 MeV/c$^{2}$) is the muon mass, $E_{BE,1s}$ is the binding energy of the $1s$ state, and $E_{recoil}$ denotes the nuclear recoil energy; $E_{\mu e}$ is nuclear dependent, for instance, $E_{\mu e}$ = 104.97 MeV for aluminum (Al). Radiative corrections have been calculated and are discussed in Ref.~\cite{szafron}. The time distribution of \MuToEmConv~ depends on the lifetime of the muonic atom ($\tau_{N}$) which again is nucleus dependent, $\tau_{N} = 0.864$ $\mu$s in aluminum.

Experiments searching for \MuToEmConv~ have the advantage of being free of accidental background events as in decay experiments such as  $\mu^{+} \rightarrow e^{+} + \gamma$ and $\mu^{+} \rightarrow e^{+} + e^{+} + e^{-}$ where free muon decays close in time can fake a signal. Experiments searching for \MuToEmConv~ are designed to be ``background free." There are two general classes of backgrounds in \MuToEmConv~ experiments:
\begin{itemize}
    \item Backgrounds that scale with the number of muons. 
    \begin{itemize}
        \item The first of these arises from muons in the 1s state that undergo normal weak decay while in orbit: $\mu \rightarrow e \nu_{\mu} \bar{\nu}_e$. The energy, $E_{\mu e}$, of the outgoing conversion electron is well above the end-point energy of the free muon decay spectrum ($\sim 52.8$ MeV) but when the decay occurs in the presence of a nucleus the outgoing electron can exchange momentum with the nucleus, resulting in a long recoil tail, parameterized in Ref. \cite{czarnecki}. Experiments searching for \MuToEmConv~ must design their tracking detectors for high momentum resolution to mitigate against these tail electrons. 
        \item Another background arises from the radiative pion capture (RPC) process, $\pi^-N \rightarrow \gamma N^{\prime}$, when a $\sim$ 105 MeV/c electron is produced from the photon pair-production, with rates as given in Refs.~\cite{bistirlich,kroll,joseph}.  Both Mu2e and COMET use a pulsed proton beam to overcome this background.  Pions have a short (26 ns) lifetime compared to muons (2.2 $\mu$s) and the experiments ``wait out" the pion decays until the background is acceptably low.  Mu2e will wait $\sim$ 700 ns in the 1695 ns period of the Fermilab Delivery Ring and COMET will use a time window from 700 to 1170 ns. Potential backgrounds from beam electrons would also scale with the number of muons.

    \end{itemize}
    \item Backgrounds that scale with running time.
    \begin{itemize}
        \item Cosmic-ray muons can knock electrons out of the stopping target or provide other sources of backgrounds which produce $\sim$ 105 MeV/c particles. Neutral secondaries from cosmic rays can also generate backgrounds. Both Mu2e and COMET employ active cosmic ray veto systems to eliminate cosmic-ray induced backgrounds.
    \end{itemize}
\end{itemize}

\subsection{Other Physics Channels}

In addition to the primary search channel, searches for \MuToEmConv~ can also have sensitivity to other CLFV BSM processes such as:

\subsubsection{ $\mu^{-} + N \rightarrow e^{+} + N'$}

In both Mu2e and COMET Phase-I, a search for $\mu^{-} + N(A,Z) \rightarrow e^{+}  + N(A,Z - 2)$, can be conducted in parallel to the primary, \MuToEmConv, search. This conversion violates both lepton number (LNV where $\Delta L =2$) and lepton flavor conservation. Searches for \MuToEpConv~ complement searches for the LNV  neutrinoless double beta decay ($0\nu\beta\beta$) process. While $0\nu\beta\beta$ looks for a ``flavor diagonal" transitions, \MuToEpConv~ looks for an ``off-diagonal" transition. Some BSM models predict flavor-diagonal transition which are suppressed with respect to those of different-flavor processes, for instance see Ref. \cite{PhysRevD.95.115010}. Measurement of a \MuToEpConv~ signal could suggest existence of non-zero Majorana neutrino mass. The ``Black Box Theorem" \cite{PhysRevD.25.2951,TAKASUGI1984372,HIRSCH2006106}, for example, relates the Majorana neutrino mass to the amplitude of any $\Delta L =2$ process and shows that the existence of such a signal implies a non-zero Majorana neutrino mass. Ref. \cite{Engel_2017} explains the LNV without introducing Majorana neutrino exchange, and instead add a new mediator particle from a super-symmetric theory or a Majoron.


The current upper-limit for \MuToEpConv~ is taken from the SINDRUM-II experiment: $Br(\mu^{-} + Ti \rightarrow e^{+} + Ca) < 1.7 \times 10^{-12} (GS) (90\% CL)$ \cite{1998334}. Mu2e and COMET could improve upon this limit by several orders of magnitude. The Mu2e sensitivity to $\mu^{-} \rightarrow e^{+}$ extends beyond the current best limit, with a $<m_{e\mu}>$ effective Majorana neutrino mass scale sensitivity down to the MeV region \cite{PhysRevC.70.065501}, surpassing the $<m_{\mu \mu}>$  sensitivity in the kaon sector which is limited to the GeV region. It should be noted that, although Mu2e and COMET Phase-I can search completely in parallel, with no dedicated run needed, COMET Phase-II would require a dedicated positron run to make this measurement. 

The experimental signature of a \MuToEpConv~ conversion process is a monochromatic $\sim$100 MeV/c (for the ground state transition (GS)) positron emanating from the stopping target; the exact momentum of the positron depends on the choice of target material. The process of \MuToEpConv~ conversion is an incoherent process, and a $\mu^{-} + N(A,Z) \rightarrow e^{+}  + N(A,Z - 2)_{GS}$ transition, where a daughter nucleus is produced in the GS, has a much better chance to be separated from the background than transitions to the excited states of the $N(Z-2,A)$ nucleus. Optimization of the experimental sensitivity has to consider the probability of transition between the ground states of the initial and final state nuclei. This probability has been calculated in Ref. \cite{Altransition} for Al-27 only. 

High energy photons from Radiative Muon Capture (RMC), $\mu^{-} + N(Z,A) \rightarrow N(Z-1,A)+\nu+\gamma(\rightarrow e^{+}e^{-})$, where the photon converts either ``externally" via in-medium pair production, or via virtually mediated ``internal" pair production, presents a dominant background to the \MuToEpConv~ searches in Mu2e and COMET. Mu2e and COMET will measure RMC spectrum. RMC is historically understudied and in order to understand any potential signals at either experiment it is important understand the shape and overall normalization of the RMC positron spectrum near the endpoint since this region overlaps with the conversion positron signal region. Most studies of nuclear RMC rely on approximations, such as the closure approximation, which are unreliable near this endpoint \cite{PhysRevC.21.1951,CHRISTILLIN1980331,MEASDAY2001243}. There have been no focused studies looking at this endpoint and there is a need for better predictions of both the real photon spectrum from RMC on medium heavy nuclei (such as Al-27), as well as the spectrum of the resultant positrons. Ref. \cite{Plestid:2020irv} presents a recent study  which describes how the near endpoint internal positron spectrum can be related to the real photon spectrum from the same nucleus, which encodes all nontrivial nuclear physics. 


\subsubsection{\MuToEX}


Experiments searching for \MuToEmConv~ could also search for a light neutral invisible particle $X$ in a bound, \MuToEX~ decay. There are a range of candidates for the particle denoted as $X$, including an axion-like particle (ALP), a familon, or a majoron with lepton flavor violating coupling to leptons.

The large number of stopped muons expected in Mu2e and COMET provide large statistics and discovery potential to probe \MuToEX~ to an unprecedented scale. From preliminary studies, an experiment with data corresponding to the experimental sensitivity of \MuToEmConv~ conversion of $\mathcal{O}(10^{-17})$, could provide sensitivity of $B(\mu \to eX) \sim \mathcal{O}(10^{-8})$. However to make a more reliable estimation of the sensitivity, several issues, such as uncertainties of the background spectrum shapes and the energy dependence of the detection acceptance, have to be carefully considered.

\clearpage

\section{DeeMe}\label{DeeMe}
\def\mue{$\mu$-$e$}
\def\FDir{deeme/figs}
\subsection{Overview}
\begin{figure}[h]
\centering
\includegraphics[width=0.5\textwidth]{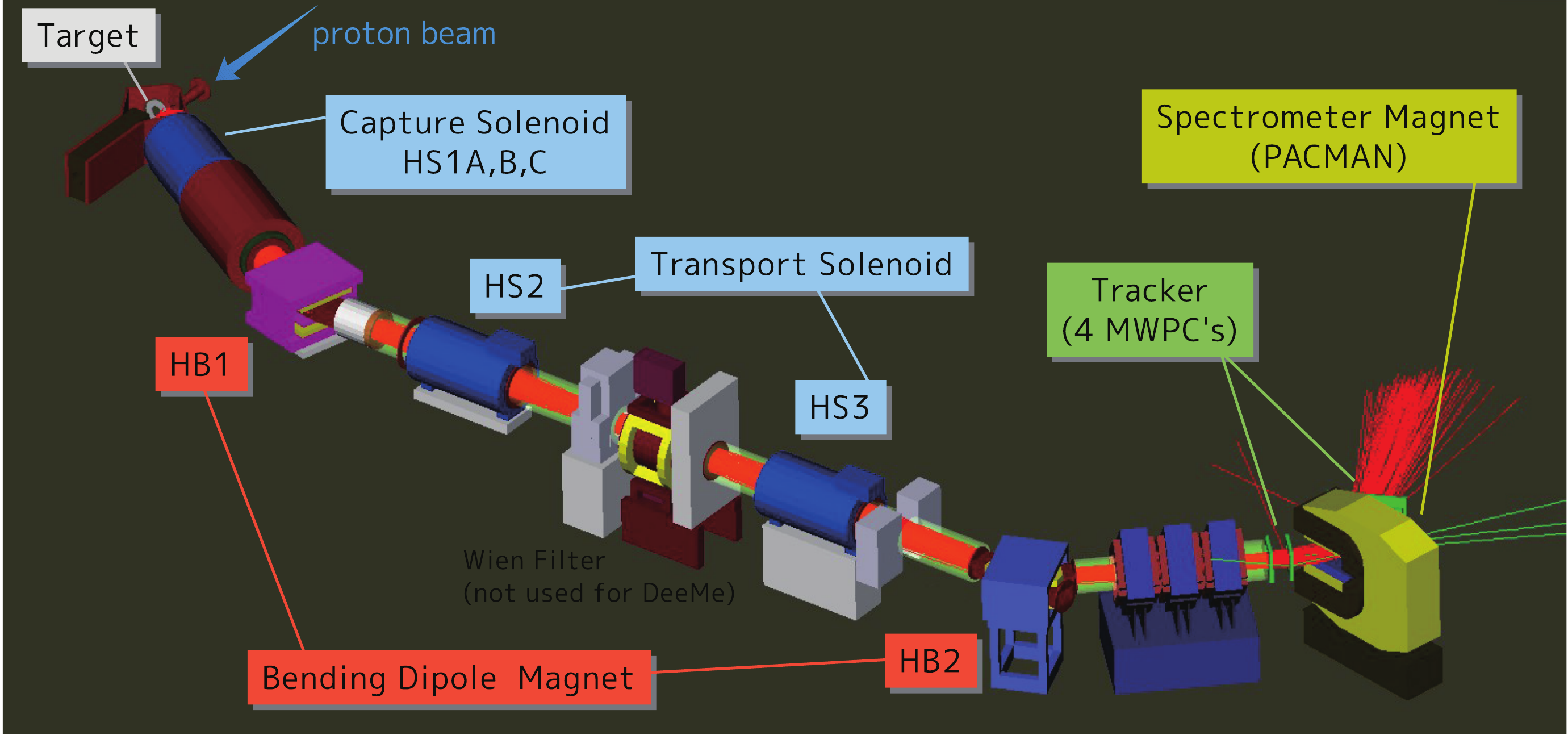}
\caption{DeeMe experimental setup.}\label{fig:overview}
\end{figure}

The DeeMe (Direct Emission of Electrons from Muon to Electron conversions)
experiment searches for neutrinoless muon to electron conversions in nuclear
fields by utilizing muonic atoms produced in a primary proton-target.
The experiment is planned to be conducted 
at the Materials and Life Science Experimental Facility (MLF)
of J-PARC.
Electrons from \MuToEmConv~ conversions
are transported by a secondary beamline to an experimental
area where 
a compact magnetic spectrometer, consisting of a dipole magnet and four sets of
multiwire proportional chambers (MWPCs), are located
to detect the electron, and measure its momentum.
An overview of the experimental setup is shown in Fig.~\ref{fig:overview}.


\subsection{Experimental apparatus}
\subsubsection{Production of muonic atoms}
Two bunches of 3 GeV protons
from the J-PARC Rapid-Cycle-Synchrotron booster (RCS)
are transported to the primary graphite target in MLF at a rate of 25 Hz.
The bunch length is 300 ns and the two bunches are separated by 600 ns.
A fraction of the low-energy pions produced by the primary proton beam
will decay in flight to muons within a few centimeters 
of their production. 
Some muons from the pion decay-in-flight will stop in the production 
target itself.
The muonic atom yield in the production target 
has been measured and is estimated to be $10^{10}$ s$^{-1}$ 
for 1-MW operation of RCS, 
which agrees well with the GEANT-4 Monte Carlo simulation.
The target material is carbon, the muon lifetime in carbon is 
about 2.0
$\mu$s; 8\% of muons will be captured
by the carbon nuclei and 92 $\%$ will decay.
The conversion signal electron energy in carbon, $E_{\mu e}$, is 105.06 MeV.

\subsubsection{H-line}
A new beamline at MLF, H-line, has been designed for DeeMe, 
which will be used to extract 105 MeV/$c$ electrons 
from the target~\cite{Hline}. 
It has a large geometrical acceptance of 110 msr.
There are a huge amount of Michel electrons and positrons from muons
in the target, but most of them are blocked by the secondary beamline. 

\subsubsection{Detector}
Even if the beamline momentum is set at 105 MeV/c, 
there will be many particles emerging out of the beamline with other momenta.
Most of them are electrons from muons decaying-in-orbit in muonic atoms
produced somewhere in the beamline.
In order to reject these off-momentum electrons, 
an electron spectrometer will
be installed at the downstream 
of beamline exit. 

\begin{figure}[h]
\centering
\includegraphics[width=0.4\textwidth]{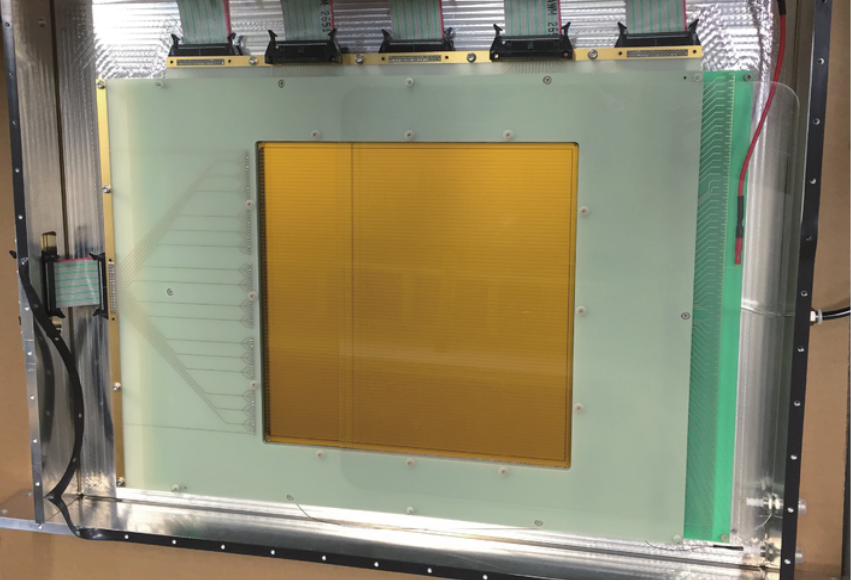}
\caption{Inside of a multiwire proportional chamber used in the DeeMe experiment.}\label{fig:mwpc}
\end{figure}

The detector system should measure the electron momentum 
with $\Delta p = 1$ MeV/$c$ of resolution 
in order to distinguish the signal from the DIO backgrounds. 
It also has to withstand the prompt charged particles of the rate that is
estimated to be $5\times 10^8/200$-nsec per pulse.
Special MWPCs with cathode strip readout have been developed 
to overcome the prompt burst~\cite{MWPC1,MWPC2}.
Anode and potential wires of 300 mm long are placed alternately with 0.7 mm 
or 0.75 mm pitch in a center plane between two cathode planes 6 mm apart.
The gas multiplication of the MWPCs is dynamically changed according to 
beam timing by switching the high-voltage applied to the chambers.
Figure~\ref{fig:mwpc} shows the inside of an MWPC.



\subsection{Physics sensitivity}
The SES to the $\mu$-$e$ conversion process is 
estimated by GEANT-4 simulation of muonic atom production in the target
and G4Beamline simulation of electron transportation through the H-line 
and the spectrometer.
The probability of muonic carbon atom production per proton on the target is
$\sim 5\times 10^{-8}$.
About 7\% of signal electrons are selected according to their momenta and 
passed to the H-line simulation, 
1\% of them go through the spectrometer simulation, 
and 70\% of them have their momenta reconstructed.
By setting the analysis time window to start 300 ns after the 2nd proton
pulse and the signal momentum region as 102.0-105.6 MeV/$c$,
about 40\% of momentum-reconstructed signal electrons remain.
The total acceptance of signal electrons is about $3\times 10^{-4}$.
The SES is expected to be 
on the order of $10^{-13}$,
where more realistic numbers depend on several factors including the RCS beam power, 
the H-line magnet power, available beam time etc..
If we consider that details of the new physics that can be probed would be
different for the different nuclei~\cite{NucleusDependence}, 
this result would be a significant achievement as the first measurement of
 \MuToEmConv~ conversion using carbon nucleus.

\subsection{Backgrounds}
For muonic carbon atoms, about 92\% of muons decay in orbit.
As described in Section~\ref{introduction} the DIO electron energy spectrum in the region lower than 52.5 MeV mostly resembles
the Michel spectrum of ordinary muon decays, 
but a high energy tail exists due to
nuclear recoils, 
and it extends up to the same energy as the \MuToEmConv conversion signal.
Thus, the electrons from DIO become one of potential background sources.
The expected DIO contribution is
estimated to be 0.03 events for the same condition that were considered for the SES evaluation.
Lower momentum regions of DIO will be used to estimate the yield of
muonic carbon atoms.

All of RCS protons are transferred to MLF by fast extraction 
and there are, in theory, no remaining protons in the RCS ring.
Also, the extraction beamline acceptance is designed to be located outside of the physical aperture of RCS and protons should hit RCS collimators 
before being accidentally extracted.
Therefore, beam-related backgrounds due to out-of-time protons are
expected to be small.
The probability of out-of-time protons hitting the target
is estimated based on actual out-of-time beam-loss measured
at the proton-extraction point of the RCS 
and the corresponding G4beamline simulation,
and found to be $3.7 \times 10^{-21}$ (90\% C.L.).
Taking this value as the normalization,
prompt electrons from the target are generated by GEANT-4, 
transported by G4beamline, 
processed with the nominal reconstruction analysis,
then the contribution of the beam-related background is estimated to be 
smaller than $10^{-4}$.

Cosmic-ray backgrounds are suppressed by a duty factor of $5 \times 10^{-5}$
as a result of 2 $\mu$s analysis window for the 25 Hz beam repetition rate.
The number of events is estimated to be less than 0.02.

\begin{figure}[h]
\centering
\includegraphics[width=0.4\textwidth]{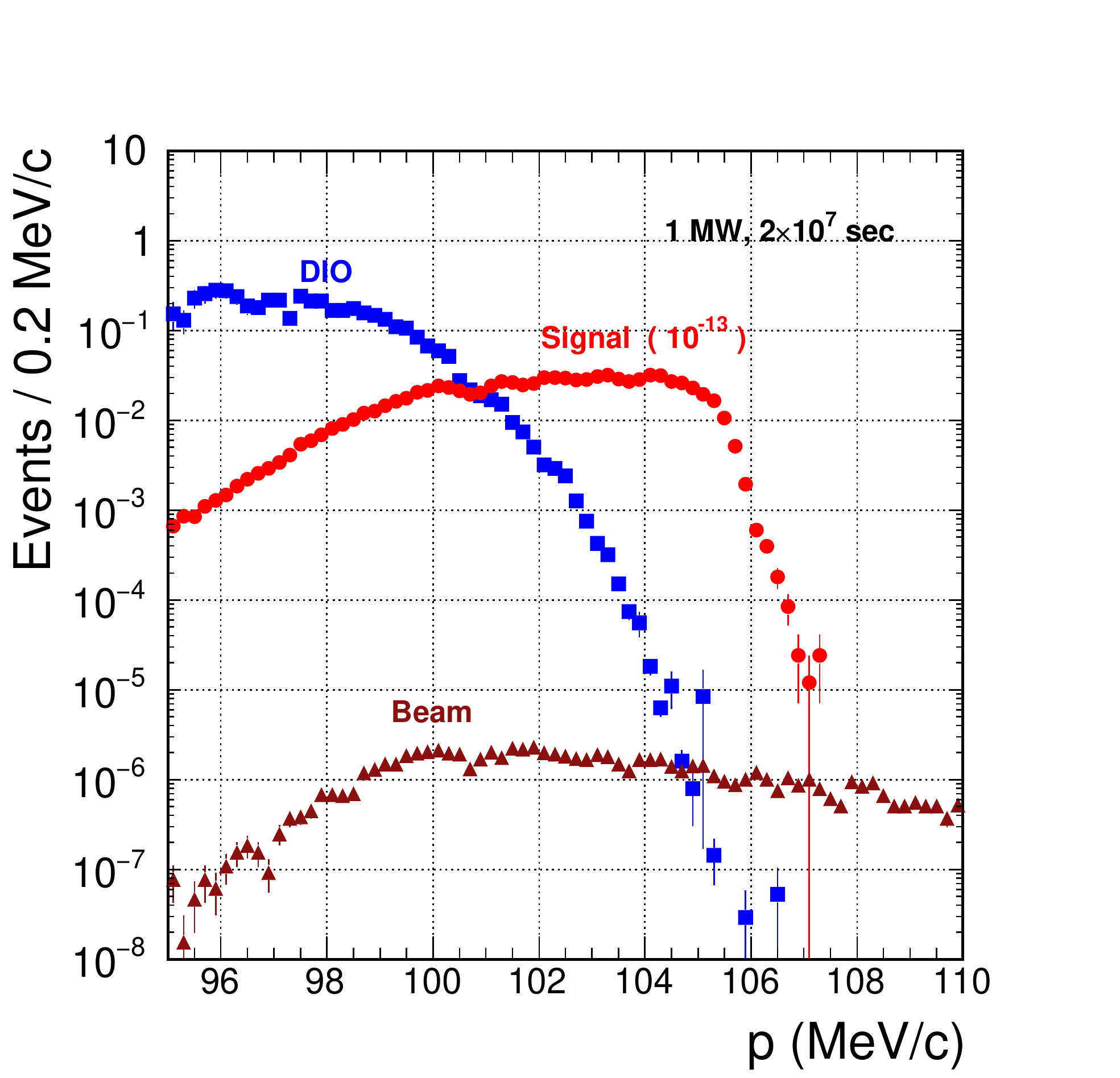}
\caption{Momentum distributions by Monte Carlo simulation
for signal electrons, DIO, and beam-related backgrounds.
}\label{fig:momentum}
\end{figure}

The expected momentum distributions for signal electrons, DIO, and 
out-of-time beam backgrounds are shown in Fig.~\ref{fig:momentum}.

\begin{figure}[t]
\centering
\includegraphics[width=0.32\textwidth]{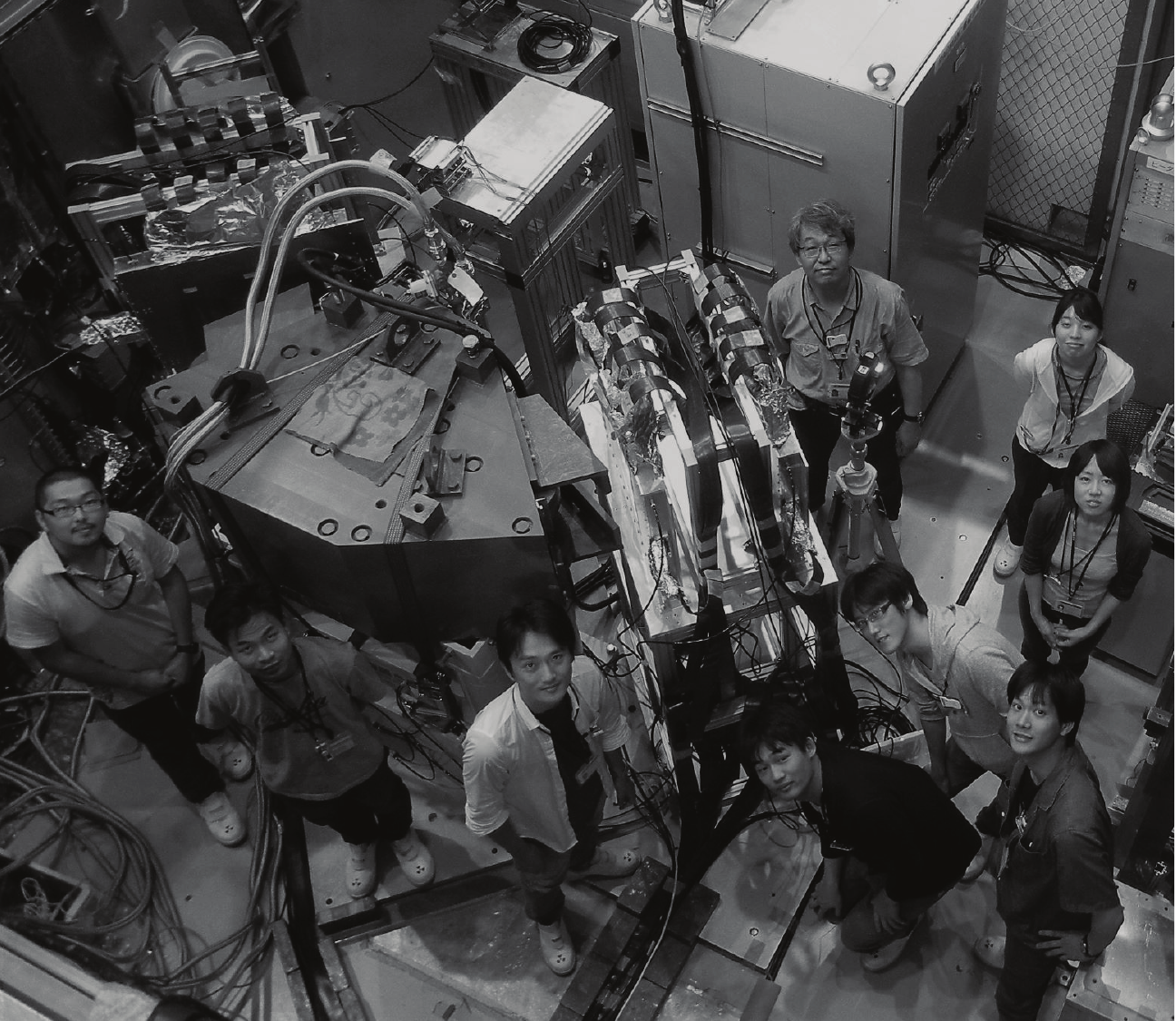}
\caption{Spectrometer setup for the DIO spectrum measurement conducted at 
the D2 area at MLF.}\label{fig:D2}
\includegraphics[width=0.4\textwidth]{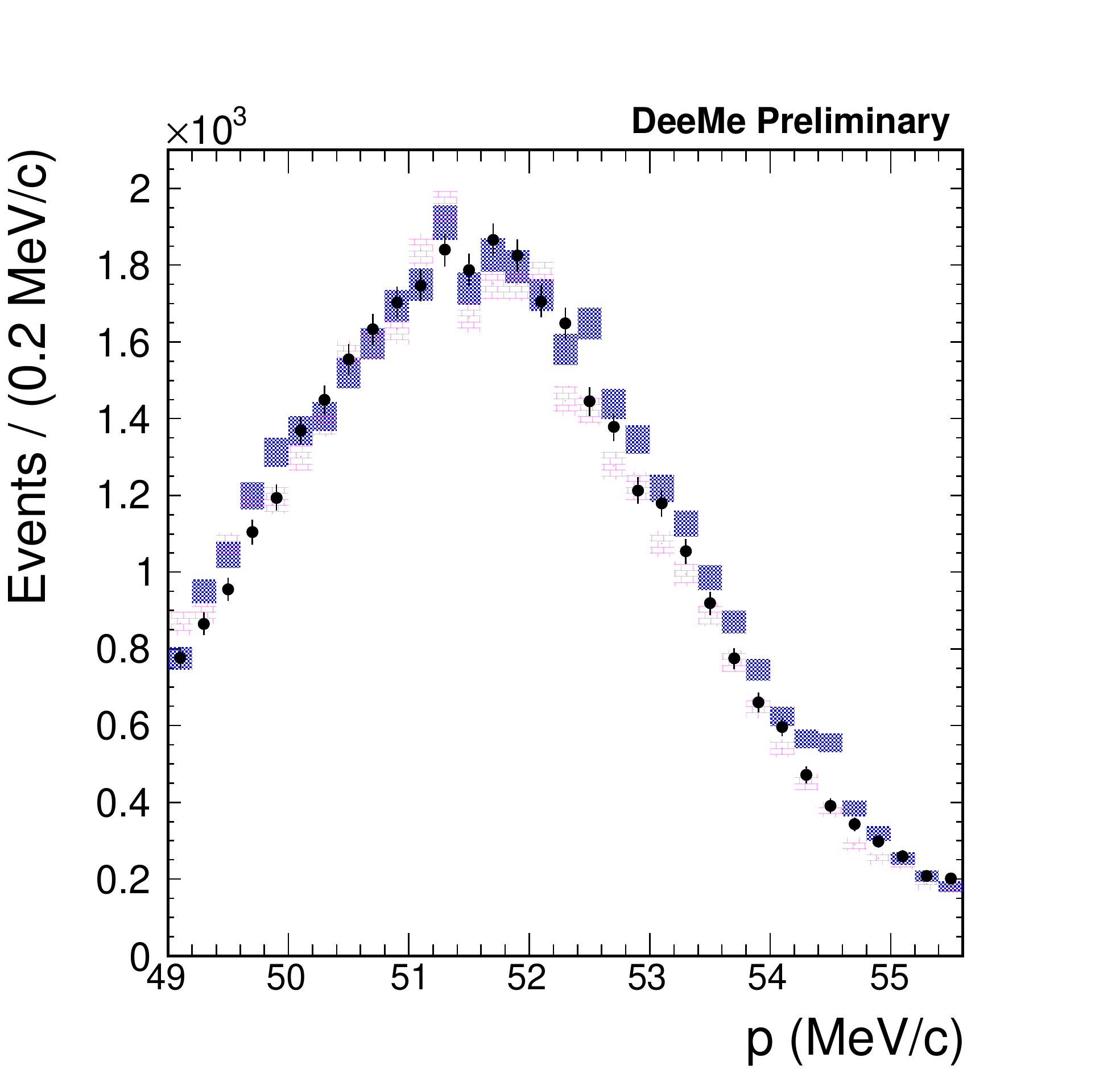}
\caption{Obtained DIO momentum distributions (points) 
compared to two kinds of theoretical calculations (shaded with two colors). A part of full data sets
was used for the data points.
}\label{fig:dio}
\end{figure}
\subsection{DIO momentum spectrum measurement}
Using an existing muon beamline at MLF, we conducted DIO spectrum measurements
with all the MWPCs and DAQ systems in place.
A graphite target was put in the D2 experimental area to stop muons.
Figure~\ref{fig:D2} is a photograph showing the setup of the spectrometer 
system.
All the systems including 384-channels of newly developed 
FADC~\cite{FADC} worked well.
Reconstructed DIO momentum spectra, obtained by using a part of full data sets, are shown in Fig.~\ref{fig:dio}.
It demonstrates that our spectrometer system is ready for beam.

\subsection{Prospects}
The H-line construction has been completed and 
it will become ready for commissioning in January 2022.
 The DeeMe collaboration plan to begin commissioning in April 2022 and
move to the initial physics runs in June 2022.
They are also expecting some beam time in the latter half of 2022
with a good level of data collection in the year 2023.

The DeeMe collaboration would like to produce physics results in a timely manner to further increase the momentum of the CLFV search activities which are expected to make unprecedented progress in the next few years.

\clearpage

\section{COMET}
\label{COMET}

\subsection{Overview}

\begin{figure}[!b]
\begin{center}
\includegraphics[width=0.6\textwidth]{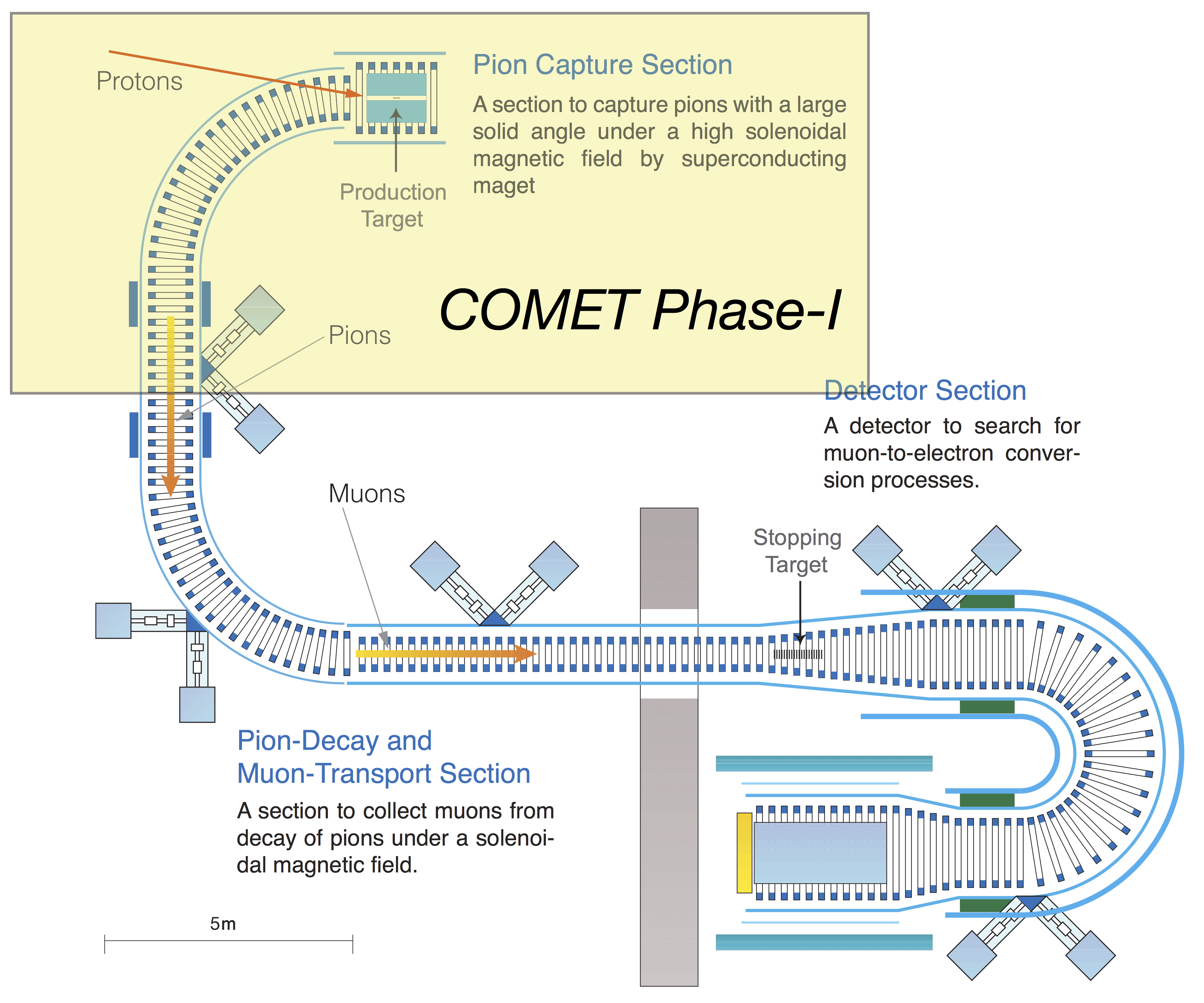}
\caption{
The COMET Phase-II experiment setup, compared with the COMET Phase-I setup.}
\label{fig:layout}
\end{center}
\end{figure}

COMET (COherent Muon to Electron Transition, J-PARC E21)  is an experimental search for the
neutrinoless conversion of a muon into an electron in the field of a nucleus
(\MuToEmConv)\,\cite{10.1093/ptep/ptz125}, 
located at the Japan Proton Accelerator Research Complex (J-PARC) in Tokai, Japan. 
The COMET experiment will be carried out in two stages. 
A half-length transport solenoid and a drift chamber detector will be used to measure a {\MuToEmConv} signal down to $\mathcal{O}(10^{-15})$ in the first stage (Phase-I) of the experiment. 
In the second stage (Phase-II)\,\cite{cometphaseIILoI,cometphaseIIthesis}, 
a full length C-shape transport solenoid, spectrometer solenoid, 
and straw tube detectors, in addition to an electromagnetic calorimeter array, 
will be used to search for {\MuToEmConv} at the level of $\mathcal{O}(10^{-17})$.
These target sensitivities are 100 or 10,000 times better than the current experimental world limit
given by the SINDRUM-II experiment. The layouts of the COMET Phase-I and Phase-II experiments are compared in Fig. \ref{fig:layout}. 

A bunched proton beam from  J-PARC is used to provide muons. Muons are captured and form muonic atoms in the 
muon stopping target which is located at the center of the cylindrical drift chamber detector of COMET Phase-I. 
The momentum of the electrons resulting from Michel decay of the muon or {\MuToEmConv} is measured by the
cylindrical drift chamber. 
In order to improve the sensitivity of the experiment, it is necessary to have
(1) a strong proton beam power to produce copious number of muons, 
(2) a long beamline for a sufficient decay of pions to muons, 
(3) a special proton bunch structure and delayed measurement timing window, and
(4) a good resolution of the electron momentum measurement.

\subsection{Accelerator and Proton beam line}
\begin{figure}[!b]
\begin{center}
\includegraphics[width=0.6\textwidth]{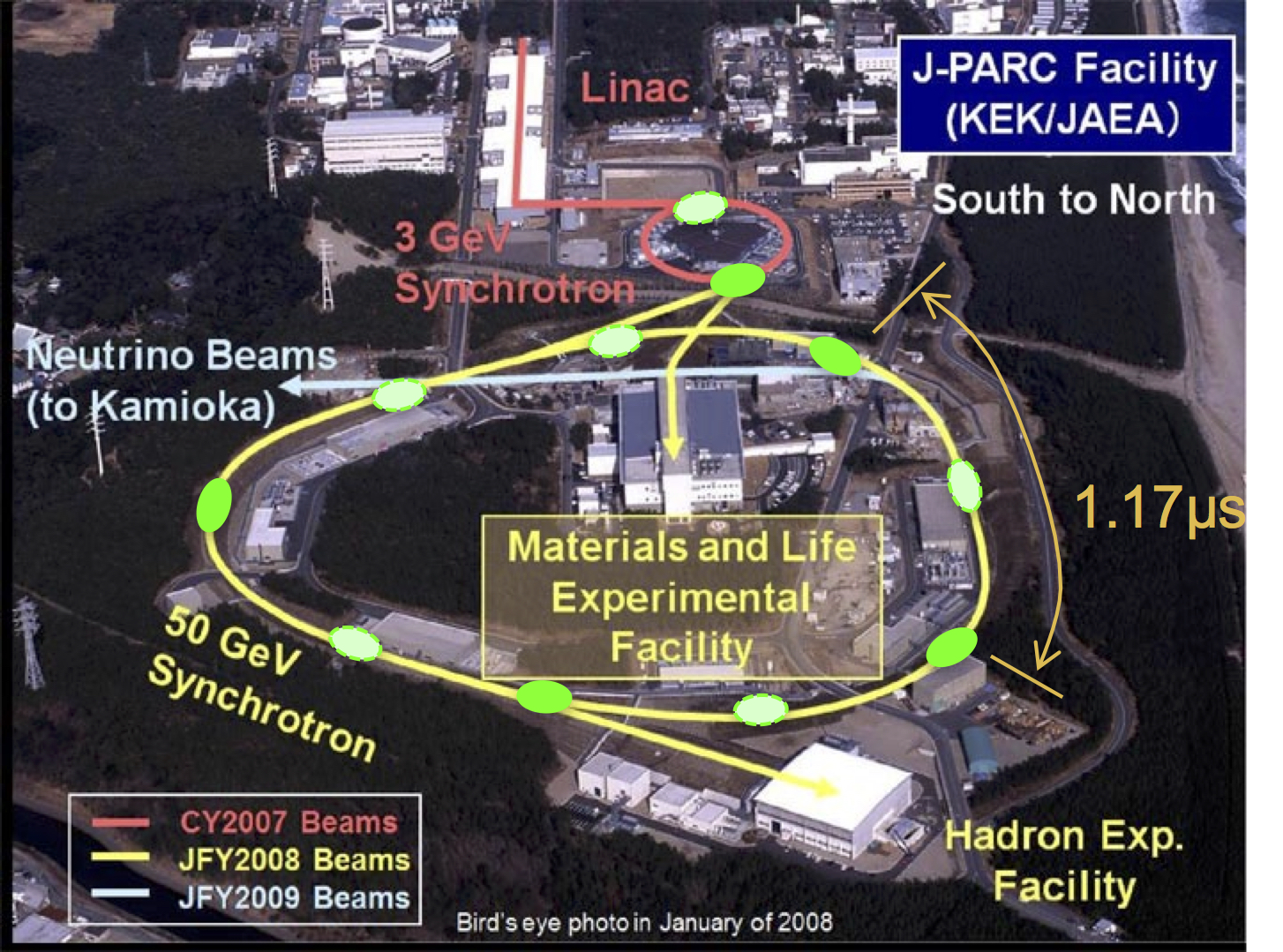}
\caption{
Bird's-eye view photo of the J-PARC facility.
The primary proton driver LINAC and 3 GeV RCS beam lines are shown in red. 
The MR and related beamlines to nuclear physics hall where the COMET experiment is located
(``Hadron experimental facility'' at the bottom right) are shown in yellow. 
The filled and hollow green circles around the MR and RCS beamlines represent the filled and empty proton buckets, respectively, showing the proton fill pattern for 1.17\,$\mu$sec bunch separation. 
Photo credit to J-PARC.}
\label{fig:comet-jparc}
\end{center}
\end{figure}

The J-PARC accelerator and proton beam lines are shown in Fig.~\ref{fig:comet-jparc}. 
The muon beam for the COMET experiment is produced from the decay of pions generated at a graphite proton target
through the interaction with 8 GeV proton beams. 
A 400\,MeV bunched proton beam generated from a linac is injected to the Rapid Cycling Synchrotron (RCS) 
and accelerated to 3\,GeV. It is again injected into the main ring (MR) of J-PARC to be accelerated to 8\,GeV. 
The designed beam power is 3.2\,kW for Phase-I and 56\,kW for Phase-II. 
The specification of the proton beam is summarized in Table \ref{tab:beam}. 

While the harmonic number of the MR is nine, only four alternating buckets are filled in order to keep the
bunch separation time at 1.17\,$\mu$sec, as shown in Fig.~\ref{fig:comet-jparc}. 
This bunch separation must be longer than the lifetime of the muonic atom 
(864\,ns for aluminum muonic atom). 
The muon decay events can be measured during the delayed time interval from the proton bunch timing,
in order to avoid a high hit rate in the detector and background hits coming from the proton pulse.
This again requires the high suppression  of the primary particles between the proton bunch, 
as the protons between the bunches coinciding with the measurement time window directly affect the {\MuToEmConv} measurement. 
The required ``extinction factor'' in the COMET experiment,
defined by the ratio of the inter-bunch particle number to the particle number in the bunch, 
is  $\mathcal{O}(10^{-10})$.
Recent efforts to improve the extinction factor of the proton beam by tuning the RCS kicker timing and MR RF voltage
have yielded  $\mathcal{O}(10^{-10})$ extinction factor in the COMET experiment\,\cite{Tomizawa:IPAC2019-WEPMP010,Nishiguchi:IPAC2019-FRXXPLS2}.

\begin{table}[tb]
\begin{center}
\caption{Proton beam specifications in the COMET Phase-I experiment.} \label{tab:beam}
\begin{tabular}{cc} \hline\hline
Beam power    & 3.2 kW \\
Energy        & 8 GeV \\
Average current & 0.4 $\mu A$ \\
Beam emittance  & 10 $\pi \rm{mm}\cdot\rm{mrad}$ \\
Proton per bunch & $ < 10^{10}$ \\
Extinction & $10^{-9}$ \\
Bunch spacing & 1.17 $\mu$sec \\
Bunch length & 100 ns \\
\hline\hline
\end{tabular}
\end{center}
\end{table}

\subsection{Pion and Muon beam line}
\begin{figure}[!b]
\begin{center}
\includegraphics[height=0.6\textwidth, angle=90]{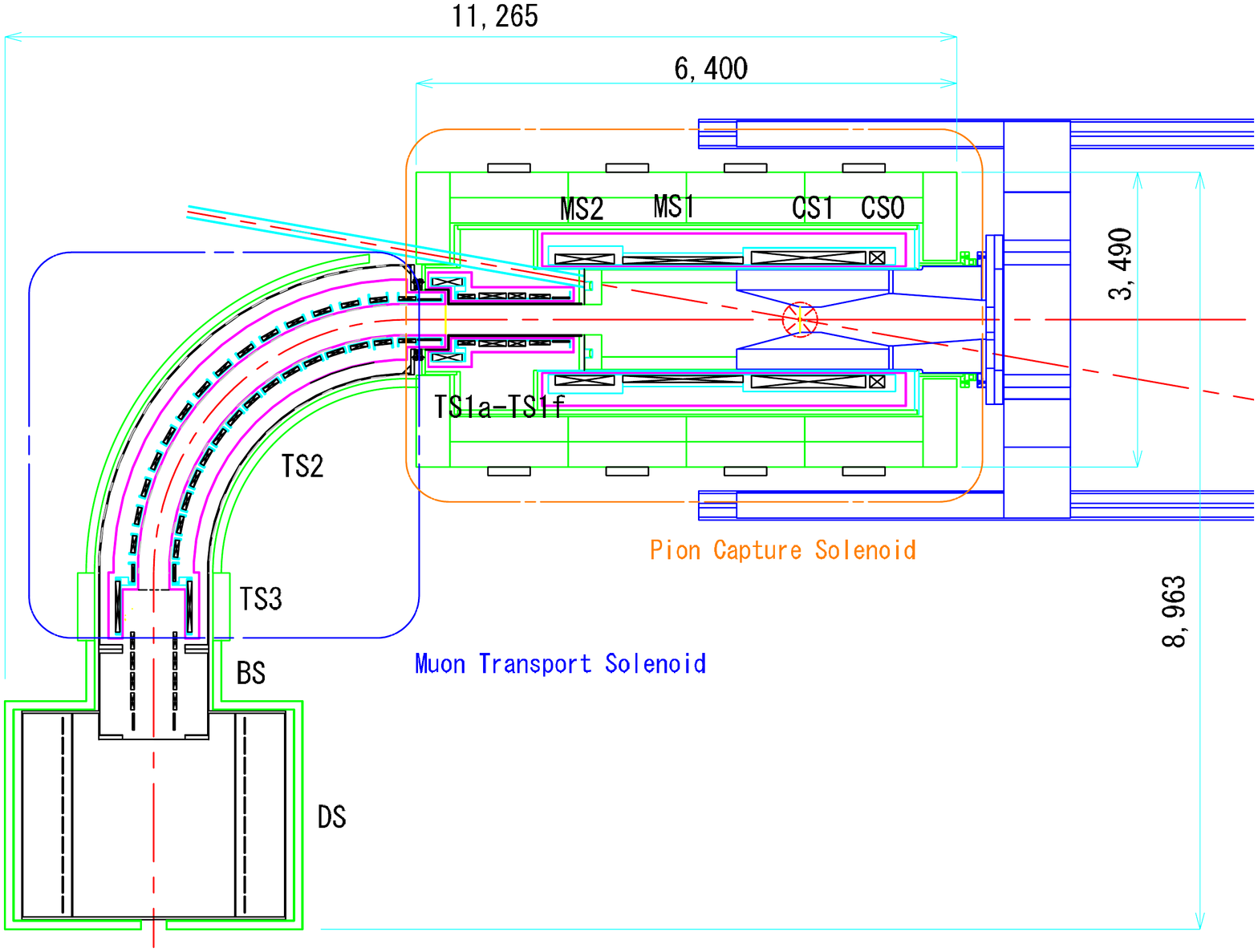}
\caption{
Layout of the COMET Phase-I solenoid system, which consists of the pion capture solenoid (``CS'', orange box), 
pion and muon transport solenoid (``TS'', blue box), 
bridge solenoid (``BS''), and 
detector solenoid (``DS'', bottom part of pion and transport solenoid). 
The slanted direction of the pion production target to the beamline is shown by the red solid-dashed line. }
\label{fig:comet-solenoid}
\end{center}
\end{figure}

The pion and muon beamline and the solenoid system of the COMET Phase-I experiment are shown in Fig.\,\ref{fig:comet-solenoid}. 
The proton target will be made of graphite or tungsten for Phase-I and Phase-II, respectively. 
The target is positioned in the center of the pion capture solenoid. The field strength values of the gradient magnetic field of the capture solenoid decreases from 5\,T at the most upstream, to 3\,T near the entrance to transport solenoid.
This ensures an adequate acceptance of backward-going pions with transverse momentum below 100\,MeV/$c$ 
to be transported into the transport solenoid. 

The pion and muon transport solenoid of Phase-II is a complete C-shaped solenoid, while 
that of Phase-I is a $90^\circ$ half length solenoid. 
The 7.6\,m-long transport solenoid for Phase-I is designed 
to optimally transport $\sim$40\,MeV/$c$ muons generated from the pion decays.  
The solenoidal field is 3\,T, and in addition to that, a vertical dipole field of $\sim$500\,Gauss is applied. 
This dipole field corrects the central axis of the helical trajectories of the charged muon  in the solenoid field
to the baseline of the solenoid,
which will otherwise vertically drift out. 
The muon and pion yields per proton at the end of the transport solenoid are estimated to 
$5.0 \times 10^{-3}$ and $3.5 \times 10^{-4}$, respectively, resulting in $\mu/\pi$ ratio of $\sim$14. 
The transport solenoid is connected to a detector solenoid through a bridge solenoid, which decreases the solenoid field 
to 1\,T. 

\subsection{The COMET Phase-I detectors}

Two separate detector systems are under construction for the COMET Phase-I experiment. In the Phase-I experiment the CyDET (cylindrical detector system) is used to search for {\MuToEmConv}.  Another detector system, StrECAL (Straw and Electromagnetic Calorimeter system), will be used in the
Phase-II experiment, The StrECAL will be commissioned during a special run period of the Phase-I experiment, for a direct beam measurement with a lower beam power.  

\begin{figure}[!b]
\begin{center}
\includegraphics[height=0.6\textwidth, angle=90]{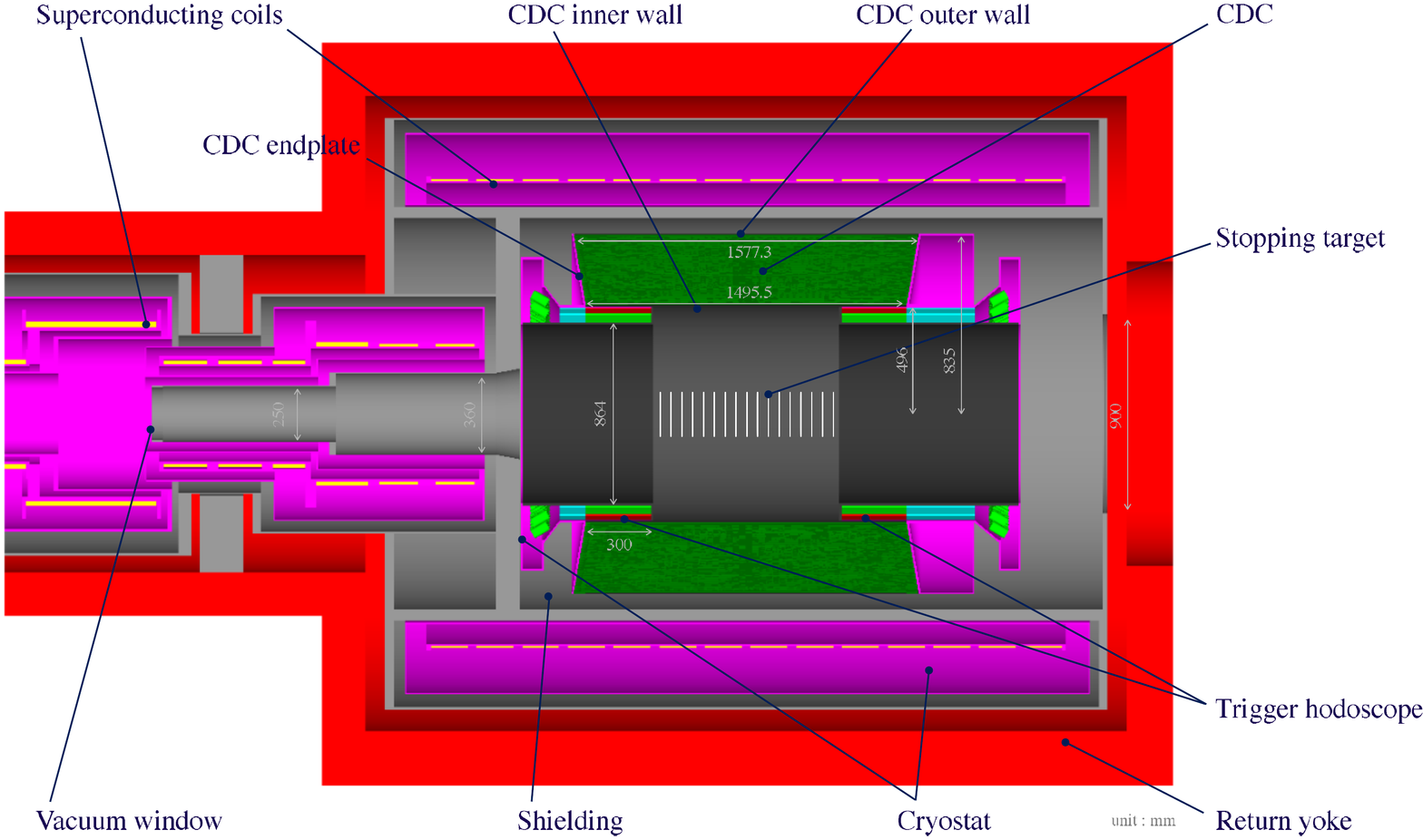}
\caption{The layout of CyDET detector system. The dark green parts are the CDC, the four red parts at CDC inner edges are the CTH, the vertical white bars at the center of CDC are the muon stopping target disks, and the magenta parts around CDC are the solenoid coils with Cryostats.} 
\label{fig:cydet}
\end{center}
\end{figure}

The CyDET, shown in Fig. \ref{fig:cydet},  is composed of  a cylindrical drift chamber (CDC) 
and two arrays of trigger hodoscope (CTH:CyDet trigger hodoscope). 
The center of the CyDET along the beam axis is designed to be hollow, with the exception of the muon stopping target. The beam particles not interacting with the muon stopping target will pass through the detector to a beam dump located at the most downstream. This design means that the CDC will remain blind to the beam particles and electrons from muon decays with a low momentum less than 60\,MeV.  
The CDC consists of around 5000 wires in 20 layers, and is filled with a mixture of helium and isobutane. From tests of a prototype, and CDC cosmic ray tests, the momentum resolution is less than 200\,keV/$c$ and the 
spatial resolution is 170\,$\mu$m \cite{WU2021165756,Moritsu:2020it,Moritsu:2019Wl}.
Two CTH rings, at both upstream and downstream edges of CDC, are composed of 64-segmented two layers of scintillators, including lead passive shielding layers, to shield from the background electrons. The CTH provides the primary trigger and event timing information 
of the {\MuToEmConv} events. The timing resolution is expected to be around 0.8\,ns. 
The muon stopping target is located in the center of the CDC, and in this target, the muons are captured
and {\MuToEmConv} can occur. The stopping target consists of 17 aluminum disks of 200\,$\mu$mm thick each. 
A separate muon stopping target monitor will be equipped to measure the characteristic 
X-ray emitted during the formation of the muonic atom, in order to directly measure the number of muons stopped. 

\begin{figure}[!b]
\begin{center}
\includegraphics[width=0.6\textwidth]{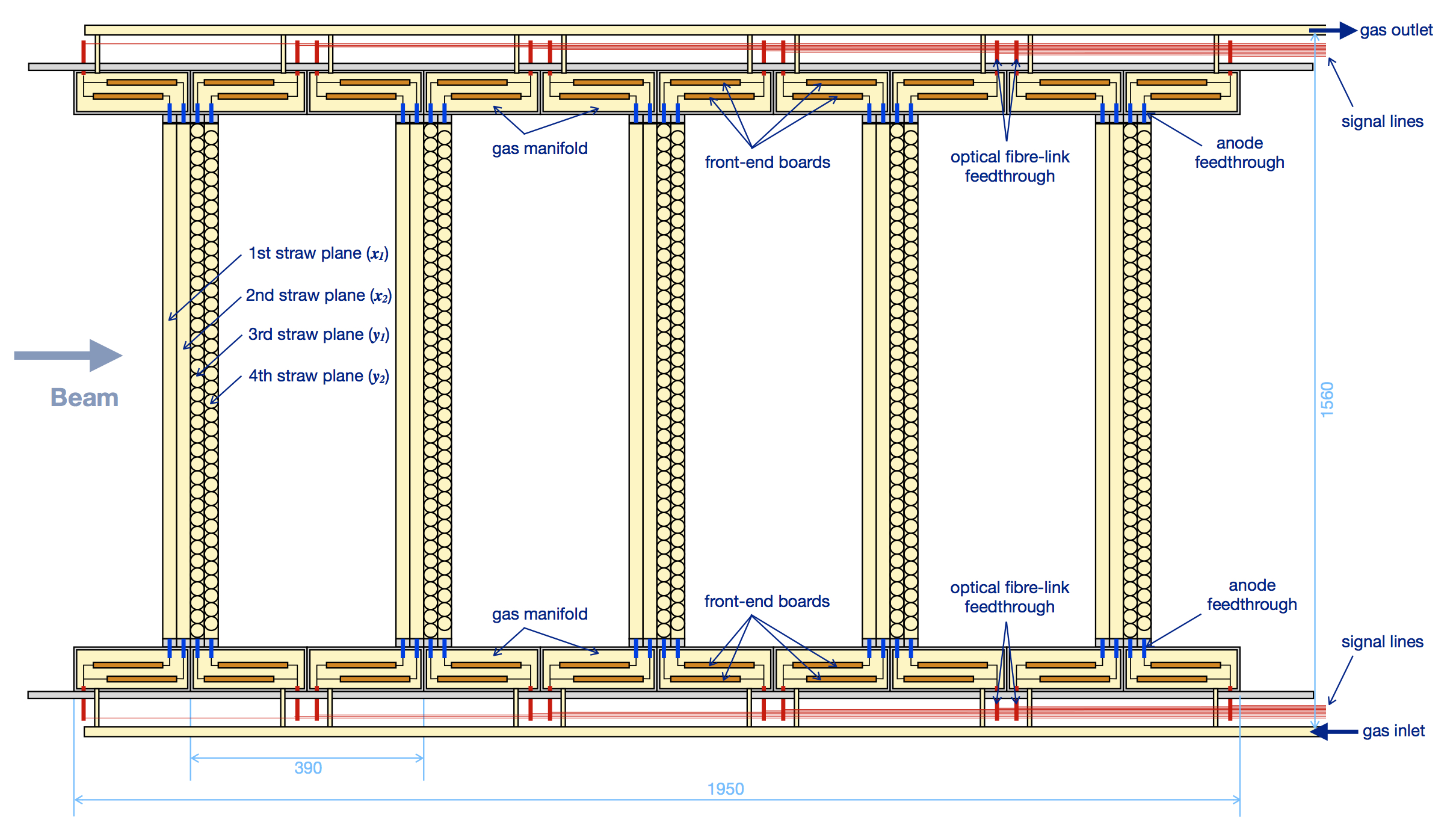}
\caption{The layout of Straw detectors of StrECAL.}
\label{fig:strecal}
\end{center}
\end{figure}

The StrECAL is composed of five straw-tube detector stations and an electromagnetic calorimeter (ECAL). 
The straw detector stations\, \cite{NISHIGUCHI2017269}, shown in Fig.\,\ref{fig:strecal}, will measure the momentum of the charged particles.
The straw tube has a diameter of 9.75\,mm with 20\,$\mu$m wall thickness, and is filled with a gas mixture of argon and ethane. 
The straws are arranged in both perpendicular directions to measure the hit position. 
From a simulation study, and several beam tests, 
the spatial resolution of the straw detector is measured to be around  150\,$\mu$m,
which is feasible to reach less than 200\,keV/$c$ momentum resolution\,\cite{VOLKOV2021165242}.
The ECAL is located at the most downstream of StrECAL, and 
is an array of $\sim$2000 LYSO (Lutetium-Yttrium Oxyorthosilicate) crystals. 
This will provide a primary trigger and also measure the particle energy. 
The StrECAL system will be used before or after  the CyDET data taking to measure the beam directly by using 1/1000 reduced beam power. This beam measurement data will play a critical role in understanding the beam-related background in both Phase-I and Phase-II experiments. 

Cosmic ray muons can decay in-flight or interact with the detector material to produce signal-like electrons.
The Cosmic Ray Veto (CRV) detector is instrumented around the CyDET (in case of Phase-I experiment) 
and the bridge solenoid, 
to identify the background induced from the cosmic rays. The 
average efficiency requirement of the CRV is 99.99\,\%. 
The CRV covering CyDET will be made of four layers of polystyrene-based organic 
scintillator strips, 
which is read out by silicon photomultipliers (SiPMs). 
In order to avoid the increase of the dark current of the SiPMs induced by the neutrons, passive neutron shieldings are 
installed inside the CRV region.
The CRV covering the bridge solenoid will be made using glass resistive plate chambers (RPC). This is 
selected due to the large flux of neutrons in the bridge solenoid area.

\subsection{Physics reach of COMET Phase-I experiment}\label{sec:physics}

The major background components in the COMET experiment are 
(1) muon decay in orbit (DIO), (2) radiative muon capture, (3) radiative pion capture, and (4) cosmic ray induced background. This high energy tail of the DIO spectrum is a non-reducible physics background source.
The momentum resolution of the tracking detector should be well understood in order to discriminate the DIO tail from the conversion electron signal. 
Radiative muon capture (RMC) occurs when the muon decays in a muonic atom through a weak interaction with the nuclei,
$\mu^- + N(A,Z) \to \nu_\mu + N(A,Z-1) + \gamma$. The photon pair-creates, producing an electron and position, $e^{-}-e^{+}$, the outgoing electron can mimic the {\MuToEmConv} signal.
In addition, the daughter nuclei is in an excited state, 
subsequently protons or neutrons are also emitted, and these are a source of background hits in the tracking detector. 
A special trigger algorithm to identify the $e^{-}-e^{+}$ pairs originating from the inner wall of the CyDET can be used to understand the
RMC background.
Radiative pion capture (RPC) is a beam-related background.
The contaminating pions in the muon beam interacts with the muon stopping target, 
$\pi^- + N(A,Z) \to N(A,Z-1) + \gamma$ or $\pi^+ + N(A,Z) \to N(A,Z+1) + \gamma$. The outgoing electron from the photon pair-creation again becomes the background. 
The primary purpose of a long muon transport solenoid and the delayed measurement to reduce pion contamination in the muon beam. 
Due to a relatively short length of the  muon transport solenoid, the COMET Phase-I experiment is subject to higher RPC background, compared to Phase II. 
Cosmic ray muons hitting the muon stopping target are an important background. In COMET Phase-I,
the cosmic ray events can be fully reconstructed in the CyDET. The CRV is also used to veto the cosmic ray induced background. 
Table \ref{tab:backgrounds} summarizes estimated background events in PhaseiI from GEANT-4 based simulation. 
The total number of background events estimated in the {\MuToEmConv}
signal region is 0.032.

\begin{table*}[tb]
\caption{Summary of the estimated background events for a
single-event sensitivity of $3 \times 10^{-15}$ in COMET Phase-I
with a proton extinction factor of $3 \times 10^{-11}$.
}
\label{tab:backgrounds}
\centering
\begin{tabular}{llr} \hline\hline
Type & Background & Estimated events \cr \hline
Physics & Muon decay in orbit & 0.01 \cr 
& Radiative muon capture & 0.0019 \cr 
& Neutron emission after muon capture & $<0.001$ \cr 
& Charged particle emission after muon capture & $<0.001$ \\ \hline
Prompt Beam 
& Radiative pion capture &  0.0028 \cr 
& Neutrons  & $ \sim 10^{-9}$ \cr 

& All others including Muon/pion decay in flight & $\le 0.0038$  \cr \hline
Delayed Beam & Beam electrons & $\sim 0$ \cr 
& Muon decay in flight  & $\sim 0$ \cr 
& Pion decay in flight  &  $\sim 0$ \cr 
& Radiative pion capture  & $\sim 0$ \cr 
& Anti-proton induced backgrounds & 0.0012 \cr \hline
Others & Cosmic rays & $<0.01$ \cr \hline
Total & & 0.032 \cr\hline\hline
\end{tabular}
\end{table*}

\begin{figure}[!bt]
\begin{center}
\includegraphics[width=0.6\textwidth]{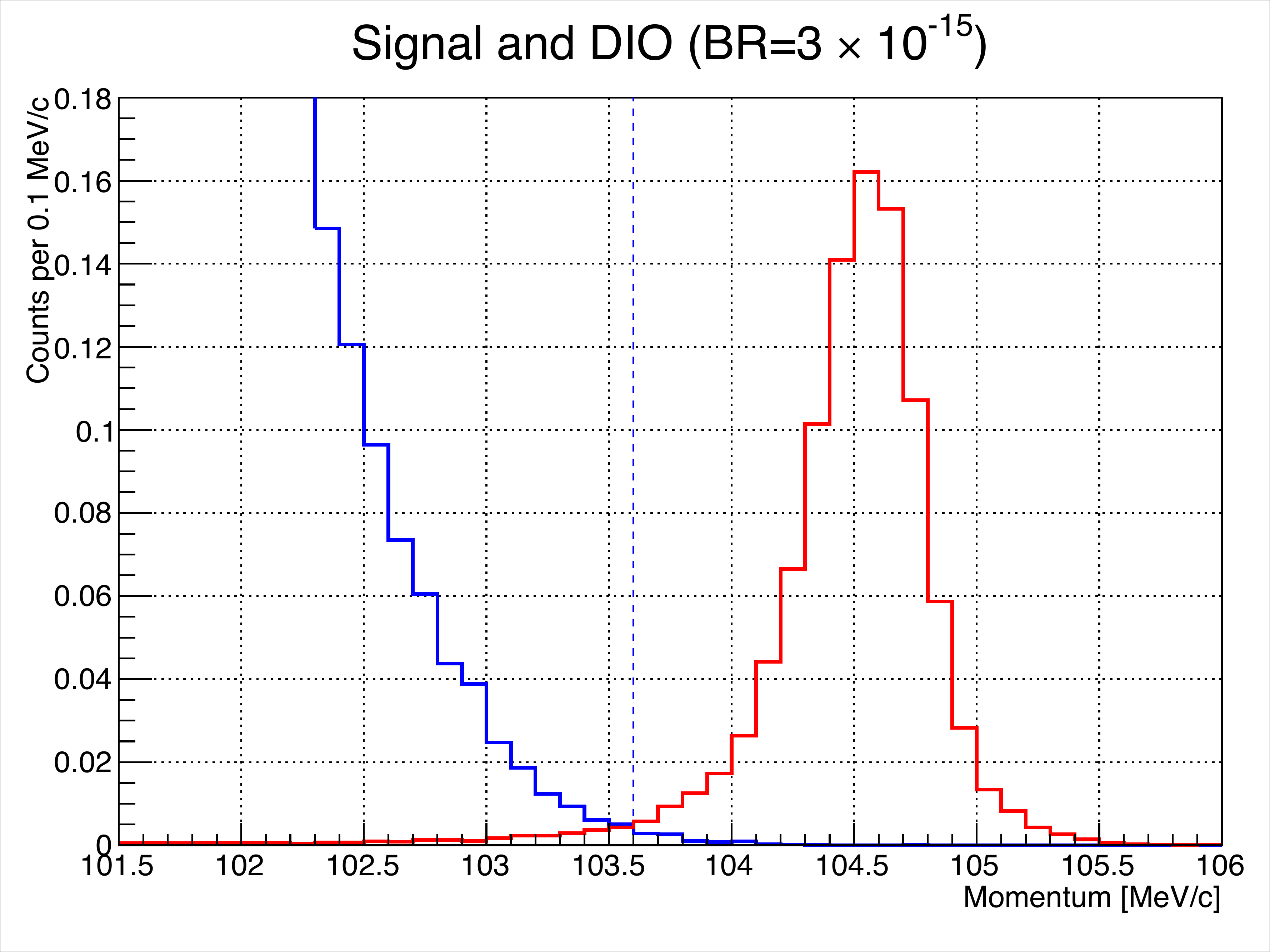}
\caption{The momentum distributions for the reconstructed {\MuToEmConv}
signals (right solid red curve with peaking distribution) 
and reconstructed DIO events (left solid blue curve with decreasing distribution).  
The vertical scale is
normalized such that the integral of the signal curve is equal to one event.
This assumes a branching ratio of $B(\mu N\rightarrow eN) = 3.1 \times 10^{-15}$.  }
\label{fig:sensitivity}
\end{center}
\end{figure}

Figure \ref{fig:sensitivity} shows the estimated DIO background spectrum and {\MuToEmConv} signal spectrum,
assuming $3 \times 10^{-15}$ conversion rate. Other backgrounds are not shown in this plot,
as their contributions are negligible. 
From a GEANT-4 based simulation framework, 
the signal acceptance ($A_{\mbox{\scriptsize $\mu$-$e$}}$) of COMET Phase-I is estimated to be 4.1\,\%.
The major inefficiencies come from 
the geometry of the CDC, which is not a $4\pi$ detector, 
and the short measurement time window due to the beam cycle. 

Based on these acceptance and background estimations, the single event sensitivity (SES) for Phase-I is:
\begin{eqnarray}
B(\mu^- + \mbox{Al} \rightarrow e^- + \mbox{Al})  &=& 
\frac{1}{N_p \cdot R_{\mu p} \cdot f_{\mbox{\scriptsize cap}} \cdot f_{\mbox{\scriptsize gnd}} \cdot A_{\mbox{\scriptsize $\mu$-$e$}}} \nonumber\\
&=& 3 \times 10^{-15} \quad {\rm (as~SES)~or } \nonumber\\
&< &7 \times 10^{-15} \quad {\rm (as~90~\%~C.L.~upper~limit)}~,
\end{eqnarray}
where $N_p = 3.2 \times 10^{19}$ is the total number of protons, 
$R_{\mu p} = 4.7 \times 10^{-4}$ is muon yield per proton obtained from simulation, 
$f_{\mbox{\scriptsize cap}} = 0.61$ is the fraction of captured muons to the total muons on target, and
$f_{\mbox{\scriptsize gnd}} = 0.9$ is the fraction of muon conversions to the ground state.

\subsection{Prospects}
The COMET experiment is a search for {\MuToEmConv}, which is one important CLFV search. 
It utilizes an 8 GeV proton beam from J-PARC to produce muons, which will be
captured in an aluminum target for muon to electron conversion. The muon conversion signal will be 
measured using a cylindrical drift chamber detector in the Phase-I experiment. 
The estimated single event sensitivity is $3 \times 10^{-15}$ which is
an improvement by a factor of 100 over the current world limit. 
The Phase-I experiment will take data over the course of 150 days beginning at the end of the Japanese fiscal year (JFY) 2023.

The detector and facility construction are on schedule. Some of important parts, such as
muon transport solenoid, detector solenoid  and drift chamber are ready and operational. 
The first straw station assembly was done in 2021, followed by assembly of additional stations. 
The J-PARC beamline for COMET is under construction during the shutdown of J-PARC MR in 2022. 
The most delayed construction is the pion production solenoid, which  is on-going, targeting to be finalized in 2023. 
The detector construction for the COMET Phase-II experiment will follow after Phase-I completion, 
targeting to achieve the single event sensitivity down to $\mathcal{O}(10^{-17} - 10^{-18})$.

\subsection{COMET Acknowledgements}
We thank KEK and J-PARC, Japan for their support of infrastructure and the operation of COMET. This work is supported in part by: the Japan Society for the Promotion of Science (JSPS) KAKENHI Grant Nos. 25000004 and 18H05231; JSPS KAKENHI Grant No. JP17H06135; the Belarusian Republican Foundation for Fundamental Research Grant F18R-006; the National Natural Science Foundation of China (NSFC) under Contract Nos. 11335009 and 11475208; the research program of the Institute of High Energy Physics (IHEP) under Contract No. Y3545111U2; the State Key Laboratory of Particle Detection and Electronics of IHEP, China, under Contract No. H929420BTD; supercomputer funding in Sun Yat-Sen University, China; the National Institute of Nuclear Physics and Particle Physics (IN2P3), France; the Shota Rustaveli National Science Foundation of Georgia (SRNSFG), grant No. DI-18-293; a Deutsche Forschungsgemeinschaft grant STO 876/7-1 of Germany; the Joint Institute for Nuclear Research (JINR), project COMET \#1134; the Institute for Basic Science (IBS) of the Republic of Korea under Project No. IBS-R017-D1-2022-a00; the Ministry of Education and Science of the Russian Federation and by the Russian Fund for Basic Research grants: 17-02-01073, 18-52-00004; the Science and Technology Facilities Council, UK; the JSPS London Short Term Predoctoral Fellowship program, a Daiwa Anglo-Japanese Foundation Small Grant; and a Royal Society International Joint Projects Grant. Crucial computing support from all partners is gratefully acknowledged, in particular from CC-IN2P3, France; GridPP, UK; and Yandex Data Factory, Russia, which also contributed expertise on machine learning methods.

\clearpage

\section{Mu2e}
\label{mu2e}

\subsection{Overview}

\begin{figure}[h!]
  \centering
  \includegraphics[width=1.0\textwidth]{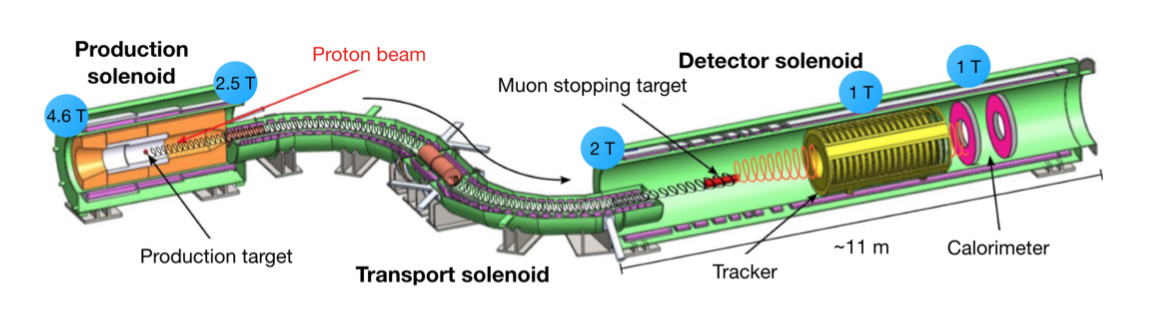}
  \caption{The Mu2e experiment consists of three super-conducting solenoids: the production (PS), transport (TS) and detector (DS) solenoids. The field is indicated. A graded field helps direct the charged particles along the desired trajectory and prevents magnetic bottles.}
  \label{fig:mu2e_layout}
\end{figure}

The Mu2e experiment is an experimental search for the coherent, neutrinoless conversion of muon into electron in a muonic atom (\MuToEmConv) currently under construction at Fermilab in the US.  Fig.~\ref{fig:mu2e_layout} gives an overview of the geometry of the Mu2e detector. Mu2e is based upon a concept proposed in Ref. \cite{MECO}, and is designed to overcome the limitations faced by SINDRUM-II:

\begin{itemize}

\item \textbf{A pulsed proton beam} is utilized, devised to produce an intense muon beam with a $\sim250$\thinspace ns wide ``microbunch" every 1695\thinspace ns. The beam is described in Section~\ref{beam};

\item \textbf{An intense muon beam} is created and transported via a series of super-conducting solenoids with graded magnetic fields providing $10^{10}\mu/s$ to the stopping target;

\item \textbf{Potential backgrounds are eliminated.} The pulsed beam helps eliminate pion backgrounds. Electrons from muon decay-in-oribt (DIO) are removed via specific tracker design choices, detailed in Section~\ref{tracker}. An active cosmic ray veto system, described in Section~\ref{CRV}, surrounds the detectors. Antiproton backgrounds are suppressed by several absorption elements installed in the muon beamline. As a result of these features Mu2e can be considered ``background-free" as only $\sim0.4$ background events are expected within the lifetime of the experiment \cite{mu2eTDR}. 

\end{itemize}

In the following sections each element in this apparatus will be described. Section~\ref{sec:timeline} will detail the current status and projected timeline of the experiment and Section~\ref{sec:physics} gives an overview of the expected physics sensitivity according to simulation.

\subsection{Accelerator and Proton beam line}
\label{beam}
\subsubsection{Pulsed Proton Beam}

\begin{figure}[ht]
  \centering
  \includegraphics[width=0.5\textwidth]{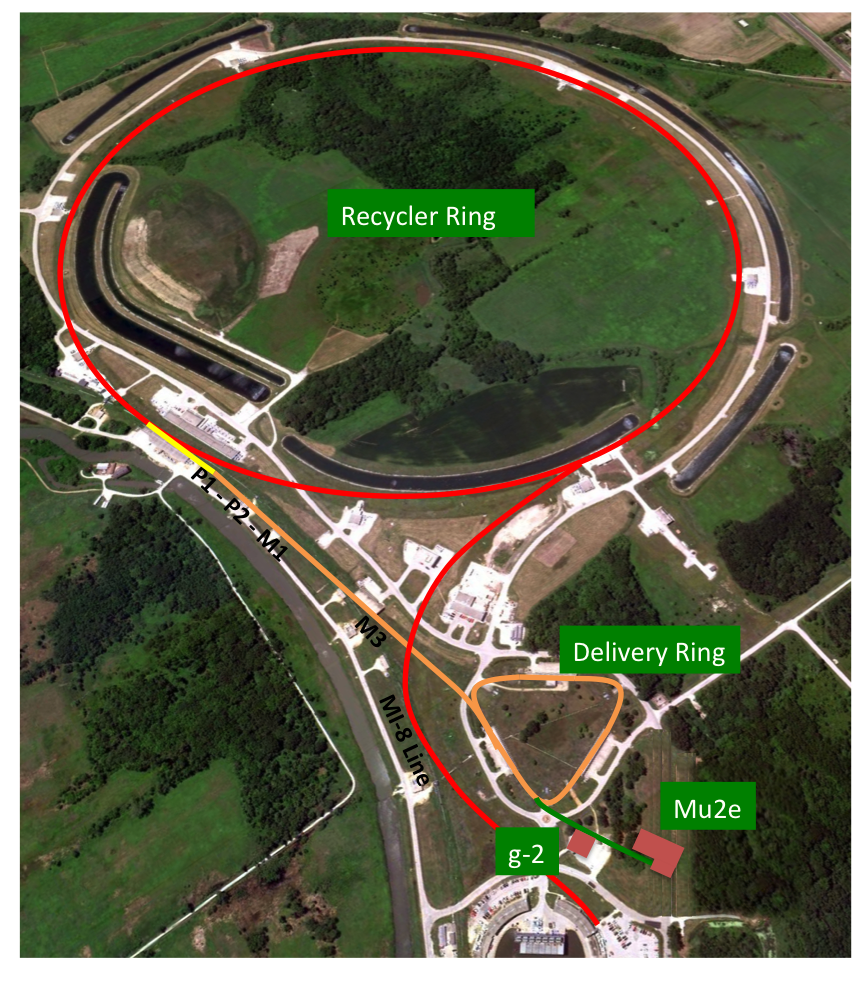}
  \caption{Layout of the Mu2e experimental hall (lower right) relative to the Fermilab accelerator complex which provides the proton beam. }
  \label{fig:FNAL_complex}
\end{figure}

Mu2e acquires its high intensity proton beam with pulsed time structure from the Fermilab accelerator complex, shown in Fig.~\ref{fig:FNAL_complex}:

\begin{enumerate}
\item Protons with kinetic energy of 8 GeV are produced in the Fermilab Booster in two batches, each of $4 \times 10^{12}$ protons. These are injected into the Recycler Ring via the MI-8 beamline;
\item A new bunch formation is performed using a RF manipulation sequence and bunches are synchronously transferred to the Delivery Ring at a rate of 2.5 MHz;
\item
  A resonant extraction system injects pulses of $\sim 3.9 \times 10^7$ protons per pulse
  into the Mu2e beam-line with a pulse spacing of 1695 ns. 
\end{enumerate}

\begin{figure}[ht]
  \centering
  \includegraphics[width=0.8\textwidth]{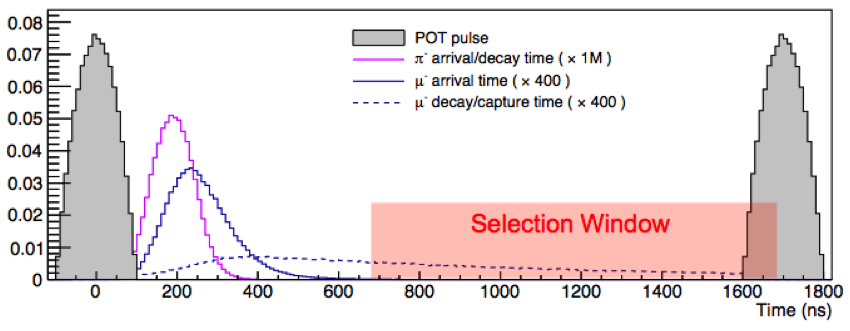}
  \caption{Proton pulses arrive at the production solenoid 1695 ns apart. A delayed livegate selection window (indicated in red) is employed in the signal search to eliminate prompt pion backgrounds which are in time with the proton pulse. }
  \label{fig:beam_struc}
\end{figure}

To reach a sensitivity of $\mathcal{O}(10^{-17})$, Mu2e must acquire a total of $\mathcal{O}( 10^{20})$ protons on target. The pulsed beam is central to Mu2e's improvement over SINDRUM-II, which used PSI's 50.1 MHz beam. The pulsed structure of the incoming proton beam can be utilized to remove pion backgrounds as pions have a much shorter lifetime than muons (26 ns at rest). As a result pion-related backgrounds decay shortly after the arrival of the proton pulse. Employing a delayed ``livegate" takes advantage of this, waiting, for example, a nominal 700 ns after the proton pulse arrives to initialize the search for the conversion signal allows efficient suppression of pion induced backgrounds, while retaining a signal acceptance of about 50$\%$. The beam structure is shown in Fig.~\ref{fig:beam_struc}. The mean lifetime of a captured muon in an aluminum nucleus is 864 ns, so the microbunch spacing is about two muon lifetimes and any conversion signal will be well-separated in time from these prompt processes. 

\subsubsection{Extinction and Monitoring}

If protons arrive at the production target between microbunches, then the resulting pion backgrounds could arrive at the same time as the muonic atoms are decaying. Such pions would not be suppressed by a delayed livegate. Consequently, to eliminate all pion backgrounds, these ``out-of-time protons" must be ``extinguished." The experiment requires a proton bunch extinction factor, defined as the ratio of out-of-time to in-time protons, of $<10^{-10}$ \cite{mu2eTDR}, while transmitting $\sim$ 99.7$\%$ of the in-time beam.  


The required extinction is achieved in two phases. Firstly, the resonant extraction process from the Delivery Ring results in an extinction factor of $10^{-5}$. In addition, the beam-line from the Delivery Ring to the production target has a set of AC oscillating dipoles that sweep out-of-time protons into a system of collimators. This results in an additional extinction factor of $10^{-7}$ or better. An extinction monitoring system, located proton downstream of the production target, provides a direct measurement of the extinction factor.

\subsection{Pion and Muon beam line}
\label{Solenoids}



Mu2e employs three superconducting solenoids to efficiently collect charged pions at the production target, and transport the negatively charged, low momentum secondary muons to the stopping target. The three superconducting solenoid systems are the production (PS), transport (TS), and detector (DS) solenoids. The PS and TS direct slow muons to the aluminum stopping target, located near the entrance of the DS. The PS is a high-field superconducting magnet with a graded field that redirects particles from proton-nucleus interactions in the tungsten production target, primarily muons and pions, towards the TS. Most pions will decay to muons in the TS. In the TS, negatively charged particles with low momentum are selected via an S-shaped solenoid configuration and several collimators. 

The Mu2e solenoids contains a continuously graded magnetic field, from 4.6 T to 1 Tesla. This gradient aids transportation and suppresses backgrounds by preventing the local trapping of particles as they traverse the muon beam-line, and by pitching beam backgrounds forward and out of the detector acceptance. The DS field has a small gradient, $\sim1.5\%$/m, to allow momentum analysis of conversion electrons. Sections~\ref{PS} - \ref{DS} outline the design and function of the three solenoids.


\subsubsection{The Production Solenoid}
\label{PS}
The PS is a high field superconducting magnet with a
graded field varying from 4.6 T to 2.5 T. The solenoid is
approximately 4 m long with an inner bore diameter of $\sim$ 1.5 m,
evacuated to $10^{-5}$ Torr. A bronze shield structure is
placed in between the inner bore and the PS coil to limit the
radiation damage.

The PS houses the Mu2e tungsten production target. Tungsten is a high-Z material and is chosen to maximize pion production. The target has a small physical profile to  minimize scattering and re-absorption of pions. 

The 8 GeV proton beam enters in the middle of the PS and strikes this radiatively cooled tungsten
target, producing mostly pions. The axially graded magnetic field
reflects the charged particles toward the low B-field region where the
PS is linked to the transport solenoid (TS). The graded axial magnetic
field creates a so-called ``magnetic bottle;" charged particles
produced at the opposite side, with respect to the TS entrance, are
reflected in the opposite direction.  This capture scheme
represents an innovative technology; according to many simulation
studies, the capture efficiency is expected to be about 1000
times larger than in conventional muon facilities~\cite{Hino2014206}.

\subsubsection{The Production Target}

\begin{figure}[ht]
  \centering
  \includegraphics[width=0.5\textwidth]{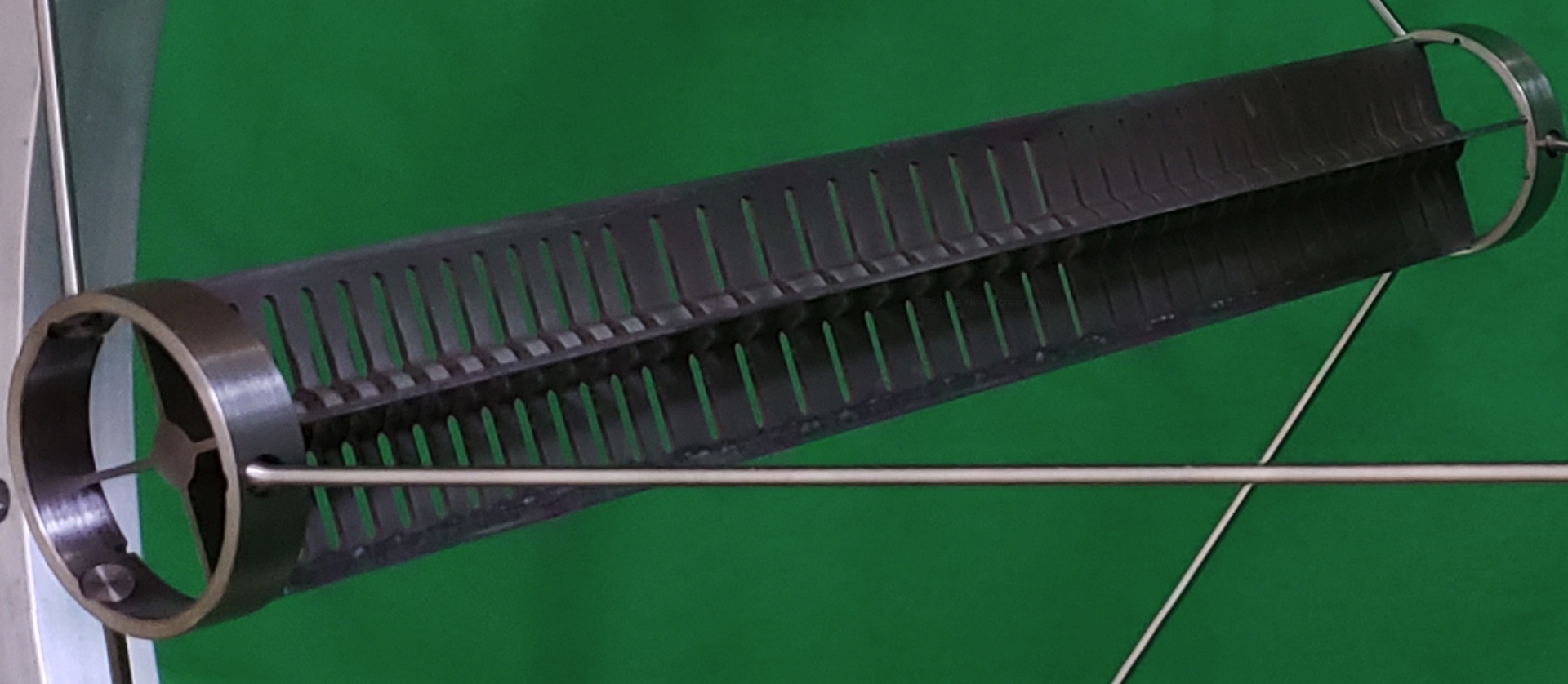}
  \caption{Image of Mu2e Production Target.  The target is in the center and the fins help dissipate radiating energy.}
  \label{fig:PT}
\end{figure}

The PS field reaches its maximum strength at about 4.6 T. To improve effectiveness of the radiative cooling, the production target design has evolved from a simple 170 mm long, 6 mm diameter cylinder to a structure with a significantly larger radiating surface. A prototype of the Mu2e production target is shown in Fig.~\ref{fig:PT}.

\subsubsection{The Transport Solenoid}
\label{TS}
The S-shaped Transport Solenoid (TS) consists of several straight and
toroidal sections. This shape helps eliminate line-of-sight backgrounds.  Collimators and absorbers, installed inside the TS, provide efficient collection and transmission of low energy, negative muons. 

\begin{figure}[ht]
  \centering
  \includegraphics[width=0.4\textwidth]{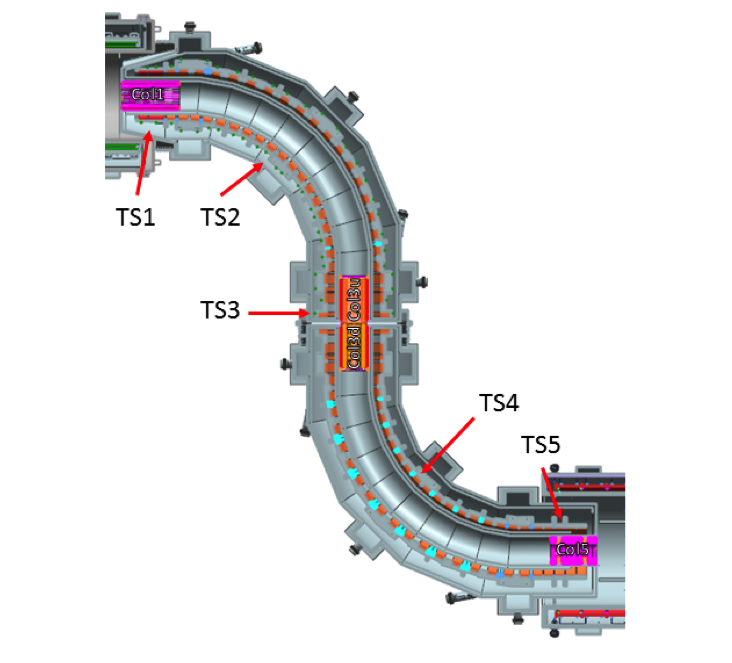}
  \caption{The Mu2e Transport Solenoid consists of five straight and toroidal sections.}
  \label{TS_view}
\end{figure}

Fig.~\ref{TS_view} shows the five main TS components. TS1 houses a collimator that selects particles with momentum $<$ 100 MeV/c. Following this, TS2, a quarter toroid, stops neutral particle propagating further. TS3 is a straight solenoid containing two collimators which filter particles based on charge and momentum, and a beryllium window stops anti-protons. TS4 is similar to TS2 and blocks remaining neutral particles. The final section, TS5, is equipped with a collimator for final momentum selection.

\begin{figure}[ht]
  \centering
  \includegraphics[width=0.5\textwidth]{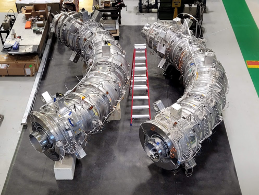}
  \caption{The Mu2e TS cold masses at Fermilab in 2020.}
  \label{fig:TS_half}
\end{figure}

\subsubsection{The Detector Solenoid}
\label{DS}

The Detector Solenoid (DS) is $\sim11$ m long and houses the muon stopping target,
tracker, and calorimeter. The inner bore is evacuated to $10^{-4}$
Torr in order to limit interactions of muons with gas
atoms~\cite{mu2eTDR}. The DS is subdivided in two regions:
\begin{enumerate}
\item an upstream side, housing the stopping target, closer to the TS, with
  a graded magnetic field ranging from 2 to 1 Tesla;
\item
  a downstream part, with an uniform 1 Tesla field, housing the
  detector system. 
\end{enumerate}

 The graded field reflects back $\sim$ 50$\%$ of conversion electrons emitted in the direction opposite the detector system.  The muon stopping target is surrounded by the neutron absorber which reduces the neutron flux resulting from muon capture events. A proton absorber, a thin polyethylene cylinder, is placed in between the stopping target and the detector system to prevent protons from reaching the detectors. Protons originate from muon nuclear captures and have energies up to a few tens of MeV \cite{MEASDAY2001243}; they are highly ionizing and can be a source of aging for the detectors.


\subsubsection{The Muon Stopping Target}
\label{ST}
The aluminum stopping target consists of thin Al foils with an optimized geometry. Around a third of muons entering the DS are stopped in this target. Stopped muons are captured in an atomic excited state, and promptly cascade to the 1$s$ state. In aluminum 61$\%$ of muons will be captured on the nucleus while 39$\%$ will decay-in-orbit. The design of the stopping target is a compromise between maximizing the fraction of incoming muons which are stopped, and minimizing the effects of energy loss and straggling on the outgoing conversion signal.

\begin{figure}[ht]
  \centering
  \includegraphics[width=0.3\textwidth]{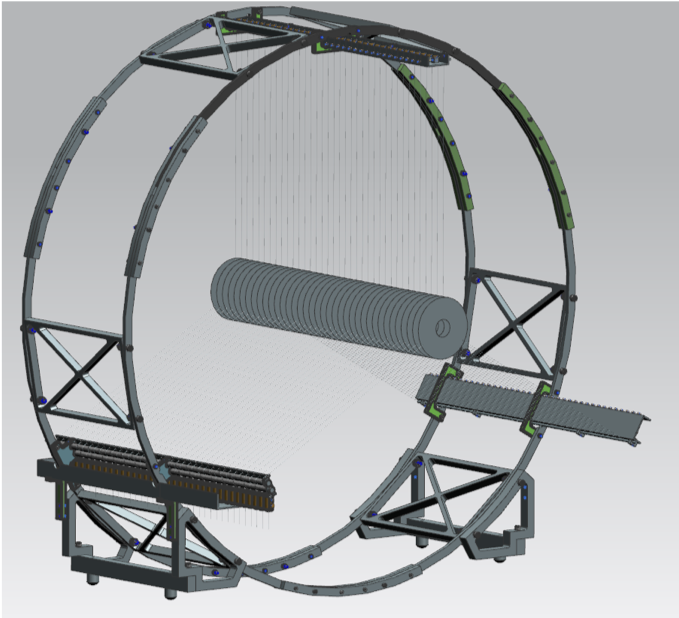}
  \includegraphics[width=0.36\textwidth]{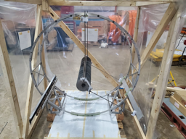}
  \caption{The aluminum stopping target designed for Mu2e consists of thin foils suspended in a frame. }
  \label{fig:ST}
\end{figure}

Fig.~\ref{fig:ST} displays the Mu2e stopping target. It consists of 37 $> 99.99 \%$ pure aluminum foils, which are 105 $\mu$m thick and 75 mm in radius, and placed 22.22 mm apart. Each foil has a 21.5 mm radius hole in the center to help reduce the effects of ``beam flash." The hole allows most beam electrons to pass through the target without interacting, reducing background radiation at the tracker. The foils are arranged co-axially along the DS axis.

\subsection{The Mu2e Detectors}

\subsubsection{The Straw Tube Tracker}
\label{tracker}
\begin{figure}[ht]
  \centering
  \includegraphics[width=0.5\textwidth]{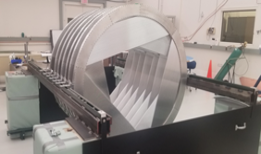}
  \caption{Some of the assembled Mu2e tracker planes in storage at Fermilab.}
  \label{fig:planes}
\end{figure}

The momentum of the electron is measured primarily through a low-mass proportional straw tube tracker. 
The tracker consists of $20,736$ straws with wall thickness of 15 $\mu$m.  The tracker is an annular cylinder and the central region is  un-instrumented. This purposely blinds the detector to nearly all muonic atom backgrounds and remnant beam, vastly reducing occupancy and making the detector insensitive to particles produced in the initial proton collision and transported to the stopping target.  With an overall geometric acceptance of about 50\%, conversion electrons will leave hits in roughly 15 straws with a reconstructed momentum resolution better than 180 KeV/c. Fig. \ref{fig:planes} shows some of the completed planes in storage at Fermilab.

\subsubsection{The Electromagnetic Calorimeter}
\label{calorimeter}

The calorimeter, downstream of the tracker, consists of two annular disks composed of $1,348$ pure cesium iodide (CsI) crystals, each coupled to two silicon photomultipliers (SiPMs). The disks have an inner (outer) radius of 35.1 (66) cm and a relative separation of 70 cm, corresponding to $\sim\frac 1 2 $ pitch of the helical conversion electron trajectory. Each disk is composed of 674 cuboid scintillating crystals of $20\ \times \ 3.4 \ \times \ 3.4$ cm$^3$. Each crystal is coupled to two $2\times3$ arrays of  $6\times6$ mm$^2$ SiPMs to read out the scintillation light. 

The calorimeter makes vital contributions to particle identification, the fast online trigger filter, precise timing information for background rejection, and to seed some instances of track reconstruction. Experimental tests with a reduced scale prototype showed that the calorimeter can provide 100 ps of timing resolution and 5$\%$ energy resolution at 100 MeV.

\subsubsection{The Cosmic Ray Veto System}
\label{CRV}
Cosmic rays induce backgrounds when they interact with material in the experiment and produce particles of $\sim$105 MeV.  This background scales with running time. This is a particular problem in the stopping target region, where a $\sim$105 MeV electron generated from a cosmic ray striking the stopping target would be indistinguishable from the conversion signal. If un-suppressed, such background events would occur at a rate of $\sim 1$ per day \cite{mu2eTDR}. Passive shielding and strict particle identification criteria at the tracker and calorimeter help remove these backgrounds. Additionally, an active veto detector, the Cosmic Ray Veto (CRV), detects penetrating cosmic ray muons. Consequently, the resulting cosmic ray
background is reduced to $\sim$ 0.10 events during the entire running period.

\begin{figure}[ht]
  \centering
  \includegraphics[width=0.45\textwidth]{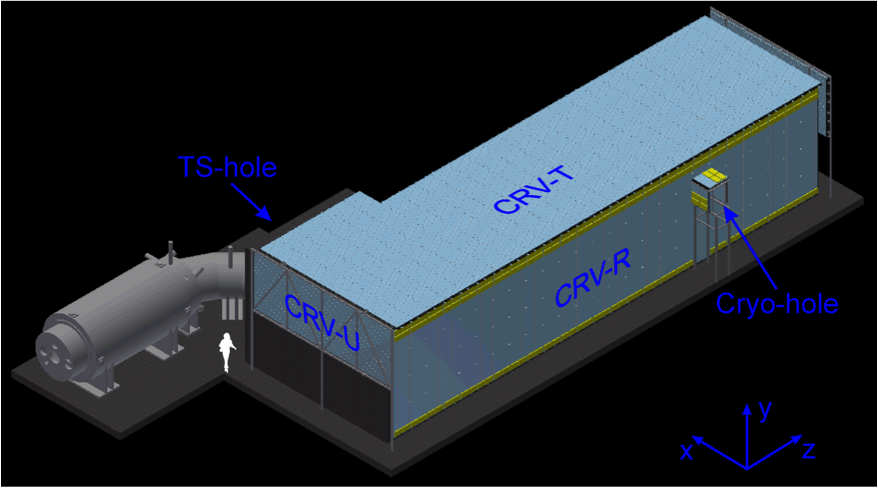}
  \caption{An active veto system surrounds the detector solenoid and downstream sections of the transport solenoid. Signal candidates are vetoed if they coincide with a cosmic event.}
  \label{fig:crv_struc}
\end{figure}

\begin{figure}[ht]
  \centering
  \includegraphics[width=0.45\textwidth]{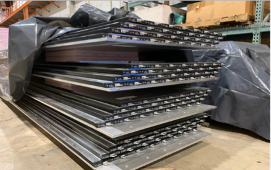}
  \caption{Some of the CRV modules in storage at Fermilab in 2021.}
  \label{fig:crv_modules}
\end{figure}

The CRV consists of four layers of extruded scintillator strips. The scintillator surrounds the top and sides of the DS and the downstream end of the TS (Fig.~\ref{fig:crv_struc}). Aluminum absorbers between the layers suppress punch through from electrons. The scintillation light is captured by embedded wavelength-shifting fibers, whose light is detected by silicon photo-multipliers (SiPMs) at each end. In the region of the muon stopping target the CRV is 99.99$\%$ efficient \cite{mu2eTDR}. Fig. \ref{fig:crv_modules} shows some of the completed modules in storage at Fermilab.

\subsubsection{The Stopping Target Monitor}

In order to precisely measure $R_{\mu e}$,  a  determination  of  the  number of muons stopped in the stopping target  is  required.  The Stopping Target Monitor (STM) will comprise of both a High Purity Germanium (HPGe) detector and a LaBr$_{3}$ detector.  The STM will count X-ray and  $\gamma$-ray emissions from muonic atoms; these provide unambiguous signals that muons have stopped in the target.

There are three unique signal lines, which occur on different time scales:
 
 \begin{itemize}
 \item Prompt: $2p\rightarrow1s$ transition produces an X-ray of 347 keV with an intensity of 79.7$\%$;
 \item Delayed: decay of activated $^{27}\text{Mg}\rightarrow$ $^{27}\text{Al}$ produces a $\gamma$-ray of 844 keV with an intensity of 71.8$\%$;
 \item Semi-prompt: the reaction $^{27}\text{Al}(\mu^- \rightarrow \nu n \gamma )$ $^{26}\text{Mg}$  produces a 1.809 MeV $\gamma$-ray with an intensity $51\pm 5\%$.
  \end{itemize}

The  peaks  for  the  first  two  lines have neighboring peaks in the vicinity.  Their measurements, therefore, require the use of a high-resolution Germanium detector. The two STM detectors will run simultaneously and provide complementary information. HPGe has a high resolution for all of the three signal peaks. In addition, LaBr$_{3}$provides poorer resolution but has a high rate capability, good for the 1809 keV peak, and does not suffer the same radiation damage which hinders HPGe. The STM detectors will be housed inside a sophisticated shield-house made out of lead, copper and aluminum, and placed downstream to the outer part of the DS to minimize the background fluxes.

\subsection{Mu2e Status and Timeline}
\label{sec:timeline}

Mu2e construction is  nearly  complete. The Mu2e beamline, which brings protons to the experimental hall, is finished. The superconducting cable for the solenoids has been procured, and the winding for all three solenoid units is well-underway. Fabrication of the TS coils is complete and all coils have now arrived at Fermilab and have been tested. Fabrication of the PS and DS coils is almost complete. 

The tracker straws and electronics prototypes have been procured. The
response of the tracker measured using cosmic rays in early production
panels meets the momentum resolution requirements. Construction of the actual tracker straws and panels is almost complete, with leak testing and plane assembly well-underway. Panel construction is $\sim$ 67 $\%$ complete, with more than 80$\%$ of straws produced. Production is scheduled to complete in 2022. All calorimeter crystals and SiPMs have been procured and the mechanical assembly of the calorimeter disks has begun. Previously, a large calorimeter prototype, containing 51 crystals and 102 SiPMs and FEE boards, was tested in an electron beam at BTF in Frascati. This demonstrated that the energy and time resolution was well within requirements and good agreement between Monte Carlo and data was again shown, \cite{osti_1523418}. The fabrication of CRV modules is almost complete, with final testing on-going. Both STM detectors have been procured and a beam test has been scheduled to allow prototyping of the data-acquisition and stopped muon counting algorithms.

Detector commissioning, without beam, will begin in 2023 and commissioning with beam will commence by the end of 2024. Physics running is expected to begin in early 2025. Physics data taking will occur in two runs separated by the two year long Fermilab accelerator shutdown which is scheduled for the end of 2026:
\begin{itemize}
    \item \textbf{Run I: 2025 - 2026 } During this time we expect to acquire enough data to provide a $\mathcal{O}(10^{3})$ sensitivity improvement on SINDRUM-II.
    \item \textbf{Run II:} Following the shutdown Mu2e will acquire the final fraction of data required to reach its design goal of a $\mathcal{O}(10^{4})$ sensitivity improvement on SINDRUM-II.
\end{itemize}

\subsection{Prospects}

\label{sec:physics}

\begin{figure}[ht]
  \centering
  \includegraphics[width=0.8\textwidth]{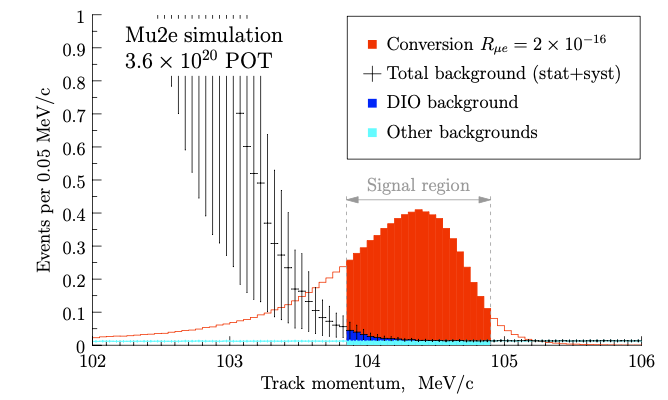}
  \caption{Momentum distributions after optimization of the signal momentum and time window.
  The background estimates are given using the optimized signal window, $103.9 < p < 105.1$ MeV/c and
  $700 < T_0 < 1695$ ns.}
  \label{fig:mu2erun1and2}
\end{figure}

Over the two run periods the total improvement in sensitivity to \MuToEmConv~ over SINDRUM-II will be $\mathcal{O}(\times 10^{4})$. Mu2e expects to reach a single event sensitivity (SES) of $\sim3.01 \times 10^{-17}$ on the conversion rate, a 90$\%$ CL upper limit on $R_{\mu e}$ of $6.12 \times 10^{-17}$, and a $5\sigma$ discovery potential at $\sim1.89 \times 10^{-16}$. These numbers, and the background yield estimates, are presented in Tab.~\ref{tbl:background_summary_optimized}, and assume $3.6 \times 10^{20}$ protons on target over the two run periods. Fig.~\ref{fig:mu2erun1and2} shows the reconstructed simulated momentum distributions of the signal and background when $R_{\mu e} = 2 \times 10^{-16}$.

\begin{table}[H]
  \centering
  \small
  \caption{
    \label{tbl:background_summary_optimized}
    Background summary and SES using the optimized signal momentum and selection window,
    $103.9 < p < 105.1$ MeV/c and $700 < T_0 < 1695$ ns.
  }
  \begin{tabular}{l|c}
    \hline
    Channel             &    Mu2e Run I+RunII                                                  \\
    \hline
    \hline
    Cosmics                  &  $ 0.209 \pm 0.022 (\text{stat.}) \pm 0.055 (\text{sys.}) $    \\
     DIO                      &  $ 0.144 \pm 0.028(\text{stat.}) \pm 0.11 (\text{sys.}) $        \\

    RPC            &  $0.021 \pm 0.001 (\text{stat.}) \pm 0.002 (\text{sys.}) $ \\
    Antiprotons              &  $ 0.040  \pm 0.001 (\text{stat.}) \pm 0.020 (\text{sys.}) $            \\
    RMC                      &  $<0.004 $          \\
    Decays in flight         &  $<0.004 $            \\
    Beam electrons           &  $0.0002 $          \\
    \hline
    Total                    &   $0.41 \pm 0.13 (\text{stat.+sys.})$                  \\
    \hline
    \hline
    SES                     & $(3.01 \pm 0.03 (\text{stat}) \pm 0.41 (\text{sys.}))\times 10^{-17}$ \\
    $R_{\mu e}$(discovery)    & $1.89\times 10^{-16}$\\
    $R_{\mu e}$(90\% CL)      & $6.12\times 10^{-17}$\\
    \hline
  \end{tabular}
\end{table}

\subsection{The Mu2e Collaboration}

In total Mu2e is made up of 38 institutions from the US, Italy, UK, China, Russia and Germany. Mu2e consists of $\sim$200 collaborators, of which 73$\%$ are at US based institutions. In addition, being based in the US, at Fermilab, of course means continued, strong US support is essential to Mu2e meeting its design goals.

\subsection{Further Ahead: Mu2e-II}

An R$\&$D effort is currently on-going to devise a Mu2e-II \cite{Mu2eII} experiment, an upgraded version of Mu2e, which would operate using the 800 MeV H- beam, PIP-II, at Fermilab. Mu2e-II will follow a similar design to Mu2e, with many components being reused. Mu2e-II is expected to reach a sensitivity of $\mathcal{O}(10^{-18})$, Mu2e-II could also be used to deploy alternative target materials. As detailed in Ref.~\cite{kitano2002} measuring the relative conversion rates in complementary materials can help elucidate the type of physics responsible for the conversion signal. 

\subsection{Mu2e Acknowledgments }

We are grateful for the vital contributions of the Fermilab staff and the technical staff of the participating institutions.  This work was supported by the US Department of Energy; the Istituto Nazionale di Fisica Nucleare, Italy; the Science and Technology Facilities Council, UK; the Ministry of Education and Science, Russian Federation; the National Science Foundation, USA; the Thousand Talents Plan, China; the Helmholtz Association, Germany; and the EU Horizon 2020 Research and Innovation Program under the Marie Sklodowska-Curie Grant Agreement No.690835. This document was prepared by members of the Mu2e Collaboration using the resources of the Fermi National Accelerator Laboratory (Fermilab), a U.S. Department of Energy, Office of Science, HEP User Facility. Fermilab is managed by Fermi Research Alliance, LLC (FRA), acting under Contract No. DE-AC02-07CH11359.


\clearpage

\section{Summary and Outlook}
High-intensity, searches for CLFV complement high energy searches for new physics.  In this paper three searches for \MuToEmConv~ have been described and their respective expected sensitivities have been detailed. All three come online in the next few years and expect to make significant improvements on the current upper limit for coherent neutrinoless muon to electron conversion relative to ordinary capture, $R_{\mu e} < 7 \times 10^{-13} $
 (90$\%$ C.L ) set by the SINDRUM-II experiment \cite{Bertl2006}. 
 
 DeeMe will begin taking data in the coming year and aims to reach a sensitivity of $\mathcal{O}(10^{-13})$ using a carbon target. COMET is a phased experiment with Phase-I, coming online in JFY 2023, having an expected sensitivity at $\mathcal{O}(10^{-15})$ and Phase II reaching $\mathcal{O}(10^{-17})$. Mu2e comes online in 2025 and aims to reach a sensitivity of $\mathcal{O}(10^{-17})$ over two run periods. Experiments such as DeeMe, COMET and Mu2e offer deep probes of new physics to an effective mass scale of up to $\mathcal{O}(10^{4}TeV/c^{2})$.

In addition, to the three experiments detailed in this paper MEG-II and Mu3e will also take data during this period. All muon CLFV experiments benefit from US participation, with Mu2e specifically having a very large US component and being based in the US. 73 $\%$ of the Mu2e collaborators are based in the US. We encourage strong support for the muon CLFV program and specifically for experimental searches for \MuToEmConv~.

\bibliographystyle{utcaps_mod}
\bibliography{main}

\end{document}